\documentclass[aps,prb,twocolumn,superscriptaddress]{revtex4-2}

\usepackage{graphicx}
\usepackage{float}
\usepackage{upgreek}
\usepackage{physics}

\makeatletter
\newcommand*{\rom}[1]{\expandafter\@slowromancap\romannumeral #1@}
\makeatother

\begin{document}


\title{Impact of coordinate frames on mode formation in twisted waveguides}




\author{Johannes Bürger}
\email{j.buerger@physik.uni-muenchen.de}
\affiliation{Chair in Hybrid Nanosystems, Nanoinstitute Munich, Ludwig-Maximilians-Universität Munich, Königinstraße 10, 80539 Munich, Germany}

\author{Adrià Canós Valero}
\affiliation{Institute of Physics, University of Graz, Universitätsplatz 5, 8010 Graz, Austria}

\author{Thomas Weiss}
\affiliation{Institute of Physics, University of Graz, Universitätsplatz 5, 8010 Graz, Austria}
\affiliation{4th Physics Institute and SCoPE, University of Stuttgart, Pfaffenwaldring 57, 70569 Stuttgart, Germany}

\author{Stefan A. Maier}
\affiliation{School of Physics and Astronomy, Monash University, Clayton, Victoria 3800, Australia}
\affiliation{Chair in Hybrid Nanosystems, Nanoinstitute Munich, Ludwig-Maximilians-Universität Munich, Königinstraße 10, 80539 Munich, Germany}
\affiliation{The Blackett Laboratory, Department of Physics, Imperial College London, London SW7 2AZ, United Kingdom}

\author{Markus A. Schmidt}
\affiliation{Leibniz Institute of Photonic Technology, Albert-Einstein-Str. 9, 07745 Jena, Germany}
\affiliation{Otto Schott Institute of Materials Research (OSIM), Friedrich-Schiller-Universität Jena, Fraunhoferstr. 6, 07743 Jena, Germany}
\affiliation{Abbe Center of Photonics and Faculty of Physics, Friedrich-Schiller-Universität Jena, Max-Wien-Platz 1, 07743 Jena, Germany}

\date{April 5, 2024}

\begin{abstract}
Off-axis twisted waveguides possess unique optical properties such as circular and orbital angular momentum (OAM) birefringence, setting them apart from their straight counterparts. Analyzing mode formation in such helical waveguides relies on the use of specific coordinate frames that follow the twist of the structure. In this study, the differences between modes forming in high-contrast off-axis twisted waveguides defined in the three most important coordinate systems - the Frenet-Serret, the helicoidal, or the Overfelt frame - are investigated through numerical simulations. We explore modal characteristics up to high twist rates (pitch: $50 \ \upmu \mathrm{m}$) and clarify a transformation allowing to map the modal fields and the effective index back to the laboratory frame.

In case the waveguide is single-mode, the fundamental modes of the three types of waveguides show significant differences in terms of birefringence, propagation loss, and polarization. Conversely, the modal characteristics of the investigated waveguides are comparable in the multimode domain.

Furthermore, our study examines the impact of twisting on spatial mode properties. At high twist rates, a separation of modes with different spin is observed, suggesting a potential influence of the photonic spin Hall effect. Additionally, twisting induces OAM-dependent changes in the intensity distribution, indicating the presence of the photonic orbital Hall effect. Lastly, modes of single-mode helical waveguides were found to exhibit superchiral fields on their surfaces.

These findings provide a comprehensive basis for further research into the physics of twisted off-axis waveguides. Implementation approaches such as 3D-nanoprinting or fiber-preform twisting open the doors to potential applications of such highly twisted waveguides, including chip-integrated devices for broadband spin- and OAM-preserving optical signal transport, as well as applications in chiral spectroscopy or nonlinear frequency conversion.
\end{abstract}


\maketitle


\section{Introduction}

The interest in off-axis twisted waveguides started emerging in the 1980s due to the observation that single-mode fibers helically coiled around a cylinder exhibit circular birefringence, manifested by the rotation of the polarization state of linearly polarized light as it propagates along the waveguide \cite{Ross1984, Varnham1985, Qian1986, Berry1987}. The rotation can be understood based on the transversality of light \cite{Ross1984}, which means that the polarization vector is constrained to the surface of the $\vb{k}$ sphere. As the wavevector $\vb{k}$ changes direction when light propagates along the helix, the polarization vector is parallel transported on the surface of the sphere \cite{Bliokh2015} - a concept well known from differential geometry \cite{ONeill2006}. Due to the curvature of this surface, it was found that the polarization vector does not necessarily return to its original state after the light completes a closed loop on the $\vb{k}$ sphere. More specifically, the polarization vector rotates relative to the laboratory frame whenever the light's trajectory features a nonzero torsion (which refers to a purely geometrical quantity and is not related to any torsional stress in the material). This conceptual framework initially served well in explaining the observed circular birefringence in helical fibers.

The theory evolved further in the years thereafter, in particular, it was realized that light traveling along curved trajectories experiences a spin-orbit and an orbit-orbit interaction coupling the spin or orbital angular momentum (OAM) of a beam of light to its orbital motion \cite{Liberman1992, Bliokh2015, Bliokh2006a, Bliokh2008}. The emergence of circular and OAM birefringence in helically coiled fibers could now be explained as a direct consequence of these couplings. Conversely, spin-orbit and orbit-orbit interactions were found to act back on the trajectory of light splitting the beam depending on its spin and OAM. These effects are now known as the photonic spin Hall \cite{Bliokh2004a, Bliokh2008} and orbital Hall effects \cite{Fedoseyev2001, Bliokh2006a}. 

From an applied perspective, helical waveguides were mostly used for their circular birefringence, which prevents linearly polarized light from becoming elliptically polarized in the presence of mechanical stress. This robustness against environmental fluctuations is employed in applications such as fiber optic current and magnetic field sensors based on the Faraday effect \cite{Rashleigh1979, Ulrich1979, Smith1978a}, or optical twist and tension sensors \cite{Ross1984}. 

In all of these early works, the cross-sectional shape of the helical waveguides is assumed to be circular in the plane perpendicular to the helical path, which can be well described in the Frenet-Serret coordinate system. On the other hand, more recent experimental works often use the helicoidal coordinate system to investigate fibers containing off-axis twisted cladding elements \cite{Wong2012, Roth2018, Ma2011a, Xi2013a, Xi2014}. However, little attention has been paid to the fact that the geometry of helical waveguides defined in these two coordinate systems is in general different which can impact their optical properties.

In this work, we present an in-depth analysis of the differences between the two coordinate frames in defining helical waveguides, and show that the resulting geometries only agree at low twist rates. Furthermore, we compare the results to helical waveguides defined in a third coordinate frame, the "Overfelt frame" \cite{Overfelt2001}, which is derived from a toroidal geometry. Note, that we always use the term "helical waveguide" to refer to an off-axis twisted waveguide, in distinction to on-axis twisted waveguides \cite{Kopp2004, Oh2004, Xu2013a, Ren2021} or two-dimensional spiraling waveguides \cite{Bauters2011, Dai2012, Lee2012}.

Another aspect frequently lacking in previous works is a comprehensive description of the transformation of fields from the helicoidal frame back to the laboratory frame. In this study, we describe this process in detail and explain under which conditions it is possible to define an effective refractive index for modes of on- and off-axis twisted waveguides in the lab frame. Such an effective index is needed for example when analyzing coupling to modes in straight waveguides or for comparison to experimental results.

We conducted simulations for both single-mode and multimode variants of the three off-axis twisted waveguide types and compared the results to an analytical model for the effective index \cite{Alexeyev2008} and loss \cite{Soh2003} of modes in helical waveguides. The study investigates twist rates of up to 20 turns per mm, which are, to our knowledge, the highest investigated so far for off-axis twisted waveguides (an overview of such works can be found in Table S\rom{1} 
of the Supplemental Material \cite{SI_Ref}). Spin- and OAM-dependent splittings in the spatial properties of the modes were analyzed, as well as the emergence of superchiral fields on the surface of the waveguides (i.e., fields with a larger chiral asymmetry than circularly polarized plane waves \cite{Tang2010,Tang2011}).

All simulations were performed for high-index contrast waveguides based on the general motivation that helical waveguides can be fabricated with such parameters using 3D-nanoprinting (two-photon-polymerization) \cite{Gao2020}.

\section{Results}

\subsection{Coordinate systems for helical waveguides}

First, we present three common coordinate systems that can be used to define waveguides that are invariant along a helical path. Throughout this work, all investigated helices are left-handed, i.e., the waveguides are "attached" to the following path $\vb{c}(z)$:
\begin{equation}
	\vb{c}(z) 
	=
	\begin{pmatrix}
		\rho \cos(\alpha z) \\
		-\rho \sin(\alpha z) \\
		z
	\end{pmatrix},
\end{equation}
where $\alpha = 2\pi/P$ is the angular twist rate, $P$ is the pitch of the helix and $\rho$ its radius. We refer to $1/P$ as the twist rate (number of turns per unit length). Such a helical path is characterized by a constant curvature $\bar{\kappa} = \rho/(\rho^2+\alpha^{-2}) > 0$ and torsion $\tau = -\alpha^{-1}/(\rho^2+\alpha^{-2})$ explained in more detail in Sec.\ S\rom{3} A 1 
of the Supplemental Material \cite{SI_Ref}. For a left-handed helix, $\alpha>0$ and $\tau < 0$. To define a helical waveguide, this one-dimensional curve needs to be extended to three dimensions. To achieve this, the first step is to establish a suitable coordinate system.
\begin{figure*}[ht] 
	\includegraphics[width=172mm]{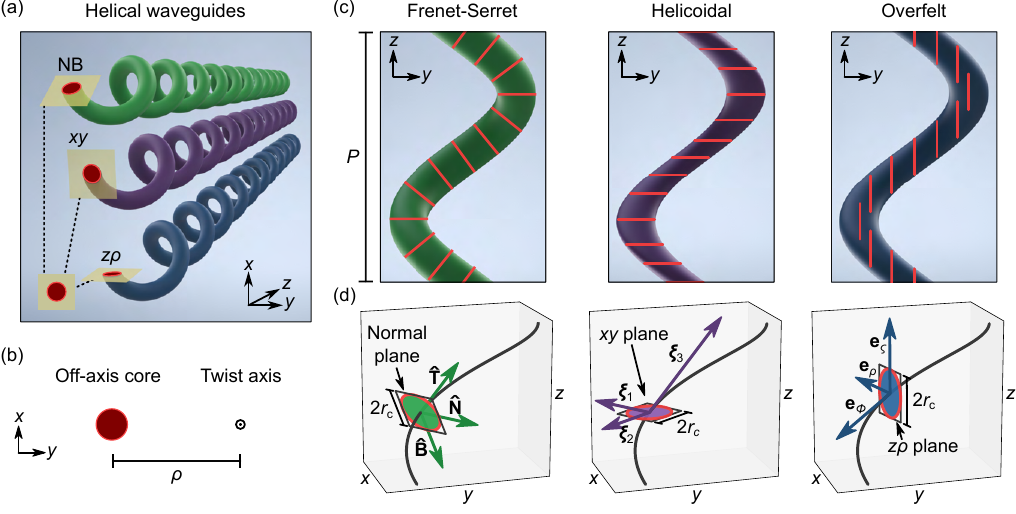}
	\caption{Helical waveguide geometries. (a) Illustration of the three waveguide types investigated in this work. (b) In these off-axis twisted waveguides, the core is located at a distance $\rho$ from the twist axis. (c) Orthographic side views of the Frenet-Serret waveguide (green), the helicoidal waveguide (purple), and the Overfelt waveguide (blue). All helices are left-handed with a pitch distance $P$. (d) Basis vectors of the corresponding coordinate systems. The waveguides are defined to have a circular cross section with radius $r_c$ in the plane spanned by the two basis vectors which are not tangential to the helical path (black curve). Red lines in (c) denote the orientation of the circular cross section. Note, that only the Frenet-Serret system is orthogonal.}
	\label{fgr:CoordinateSystems}
\end{figure*}

\subsubsection{Frenet-Serret frame}

The most natural choice for a coordinate system given any curve $\vb{c}(z)$ is the Frenet-Serret frame, a local orthonormal right-handed coordinate system that is attached to the curve. Its unit vectors are the tangent vector $\vb{\hat{T}}$ of the curve, the normal vector $\vb{\hat{N}}$ pointing along the derivative of the tangent, and the binormal vector $\vb{\hat{B}}$ which is orthogonal to $\vb{\hat{T}}$ and $\vb{\hat{N}}$ \cite{Kuhnel2015}. For the left-handed helix above these unit vectors are:
\begin{gather}
	\vb{\hat{T}}
	=
	\frac{1}{\sqrt{1+(\alpha \rho )^2}}
	\begin{pmatrix}
		-\alpha \rho  \sin(\alpha z) \\
		-\alpha \rho  \cos(\alpha z) \\
		1
	\end{pmatrix}, \;
	\vb{\hat{N}} 	=
	\begin{pmatrix}
		-\cos(\alpha z) \\
		\sin(\alpha z) \\
		0
	\end{pmatrix}, \nonumber\\
	\vb{\hat{B}} 	=
	\frac{1}{\sqrt{1+(\alpha \rho )^2}}
	\begin{pmatrix}
		-  \sin(\alpha z) \\
		-  \cos(\alpha z) \\
		-\alpha \rho
	\end{pmatrix}.
\end{gather}
In this form, $z$ is the only coordinate, while extensions of the Frenet-Serret frame for a ring of helical waveguides with an additional azimuthal coordinate exist \cite{Chen2021}. The Frenet-Serret frame will prove to be the best choice to describe off-axis twisted waveguides since the wavefronts of the modes lie within the NB plane - as revealed later in Sec.\ \ref{MultimodeOptProperties} ff. The simplest helical waveguide that can be defined in this coordinate system has a circular cross section of radius $r_c$ in the NB plane and is extended infinitely along the T direction. We refer to such a waveguide as a "Frenet-Serret waveguide", with related content being depicted in green throughout the study.

\subsubsection{Helicoidal frame}

The helicoidal frame is another coordinate system widely used to describe twisted waveguides. Helicoidal coordinates $(\xi_1,\xi_2,\xi_3)$ are related to the Cartesian coordinates $(x,y,z)$ via \cite{Russell2017}:
\begin{eqnarray}    
	\vb{r} = (x,y,z) = \;
	&&(\xi_1 \cos(\alpha \xi_3) + \xi_2 \sin(\alpha \xi_3) , \nonumber\\
	&&-\xi_1 \sin(\alpha \xi_3) + \xi_2 \cos(\alpha \xi_3) ,
	\xi_3).
\end{eqnarray}
For fixed values $\xi_1$ and $\xi_2$, the curve $\vb{r}(\xi_3)$ is a left-handed helix for $\alpha > 0$. The basis vectors $\vb{\xi_1}$ and $\vb{\xi_2}$ of the helicoidal frame lie in the $xy$ plane while $\vb{\xi_3}$ points along the tangent of the helix:
\begin{gather}
		\vb{\xi_1} =
		\frac{\partial \vb{r}}{\partial \xi_1} =
		\begin{pmatrix}
			\cos(\alpha \xi_3) \\
			- \sin(\alpha \xi_3) \\
			0
		\end{pmatrix}, \;
		\vb{\xi_2} =
		\frac{\partial \vb{r}}{\partial \xi_2} =
		\begin{pmatrix}
			\sin(\alpha \xi_3) \\
			\cos(\alpha \xi_3) \\
			0
		\end{pmatrix},  \nonumber\\
		\vb{\xi_3} =
		\frac{\partial \vb{r}}{\partial \xi_3} = 
		\begin{pmatrix}
			-\xi_1 \alpha \sin(\alpha \xi_3) + \xi_2 \alpha \cos(\alpha \xi_3) \\
			-\xi_1 \alpha \cos(\alpha \xi_3) - \xi_2 \alpha \sin(\alpha \xi_3) \\
			1
		\end{pmatrix}. 
\end{gather}
Therefore, the helicoidal coordinate system is not orthogonal. Similar to before, we define a "helicoidal waveguide" as having a circular cross section of radius $r_c$ in the $\xi_1 \xi_2$ plane and extending infinitely along $\xi_3$. Results relating to helicoidal waveguides are depicted in purple. Note, that the helicoidal coordinate system is especially useful if the wavefronts are perpendicular to the $z$ axis, which usually applies to on-axis twisted waveguides \cite{Roth2018, Xi2014}.

\subsubsection{Overfelt frame}

A third coordinate system in which a helical structure is translationally invariant along one axis is the coordinate frame $(\rho, \phi, \zeta)$ used by Overfelt, here referred to as Overfelt frame \cite{Overfelt2001}. It is an extension of the toroidal coordinate system along the $z$ axis with its coordinates being defined as:
\begin{equation}
	\vb{r}= (x,y,z) = (
	\rho \cos(\phi),
	\rho \sin(\phi),
	\zeta + \phi/\alpha).
\end{equation}
When fixing a particular value of $\rho$ and $\zeta$, $\vb{r}(\phi)$ describes a left-handed helix for $\alpha < 0$. To construct a helical waveguide, a refractive index profile $n$ is defined in the $\rho \zeta$ plane (which is identical to the $\rho z$ plane of a cylindrical coordinate system) and extended infinitely along the $\phi$ coordinate. When this profile is a circle of radius $r_c$, we refer to the resulting structure as an "Overfelt waveguide", depicted in blue throughout this work. Note, that the Overfelt system is not orthogonal:
\begin{figure*}[ht]
	\includegraphics[width=155mm ]{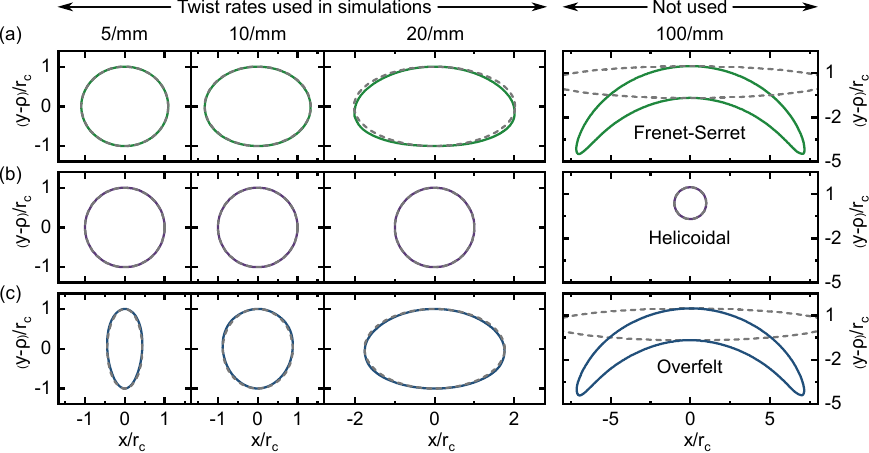}
	\caption{Cross sections of the helical waveguides in the $xy$ plane. Cross sections were calculated numerically for four different twist rates (noted on the top) using an off-axis distance of $\rho = 14 \ \upmu \mathrm{m}$ and $r_c = 1.8 \ \upmu \mathrm{m}$. The center of the twist axis is located below the shown coordinate range at $x=y=0$. The geometry of the Frenet-Serret waveguide (green) approaches that of the helicoidal waveguide (purple) for low twist rates and that of the Overfelt waveguide (blue) for high twist rates. Gray dashed lines show an analytical approximation of the cross section, which is valid in the limit of weak coiling $(\bar{\kappa} r_c \ll 1)$ and if the extent of the ellipse in $x$ direction is much smaller than $2 \pi \rho$. Waveguide modes were simulated for twist rates up to 20/mm.}
	\label{fgr:CrossSections}
\end{figure*}
\begin{gather}
	\vb{e_\rho} =
	\frac{\partial \vb{r}}{\partial \rho} =
	\begin{pmatrix}
		\cos(\phi) \\
		\sin(\phi) \\
		0
	\end{pmatrix}, \;
	\vb{e_\phi} =
	\frac{\partial \vb{r}}{\partial \phi} = \frac{1}{\alpha}
	\begin{pmatrix}
		- \alpha \rho \sin(\phi) \\
		\alpha \rho \cos(\phi) \\
		1
	\end{pmatrix}, \nonumber\\
	\vb{e_\zeta} =
	\frac{\partial \vb{r}}{\partial \zeta} =
	\begin{pmatrix}
		0 \\
		0 \\
		1
	\end{pmatrix}.
\end{gather}
More detailed descriptions of the coordinate systems can be found in Sec.\ S\rom{3} 
of the Supplemental Material \cite{SI_Ref}. Their basis vectors and the corresponding waveguides are shown in Fig.\ \ref{fgr:CoordinateSystems}. \\

\subsection{Waveguide cross sections in xy plane}

To compare the optical properties of the three helical waveguides, mode simulations are conducted using a commercial FEM solver (PropagatingMode module of JCMwave). JCMwave offers built-in support for the helicoidal coordinate system, along with appropriately defined perfectly matched layers. Application of the helicoidal coordinate system requires knowledge of the cross sections of the waveguides in the $xy$ plane, which are shown for different twist rates in Fig.\ \ref{fgr:CrossSections}. In the limit of weak coiling, i.e., if the radius of curvature of the helix is much larger than the radius of the core $(\bar{\kappa} r_c \ll 1)$, the waveguide can be "unrolled" onto a plane allowing the cross sections to be described as ellipses (gray dashed lines in Fig.\ \ref{fgr:CrossSections}). This process is explained in more detail in Sec.\ S\rom{4} 
of the Supplemental Material \cite{SI_Ref}. For the reader's convenience, the formulas for the semi-axes $r_c^x$ and $r_c^y$ of the resulting ellipses are shown in Table \ref{ParamsEllipsesMain}. 
\begin{table}[H]
	\caption{\label{ParamsEllipsesMain} Semi-major and semi-minor axes of the elliptical $xy$ cross sections of the three helical waveguides in the limit of weak-coiling $(\bar{\kappa} r_c \ll 1)$ and $2 r_c^x \ll 2 \pi \rho$.}
	\centering	
 \begin{ruledtabular}	
	\begin{tabular}{cccc}
		\rule{0pt}{2ex}
		 Semi-axis   &  Frenet-Serret   &  Helicoidal  &  Overfelt  \\[1mm]
		\hline
		\rule{0pt}{3ex}
		$r_c^x$   & $\sqrt{1+(\alpha \rho)^2} \, r_c $  & $r_c$ & $\alpha \rho \, r_c $ \\
		$r_c^y$   & $r_c$ & $r_c$ & $r_c$ \\
	\end{tabular}
\end{ruledtabular}
\end{table}

With the parameters used in this work, the weak coiling approximation is valid for all investigated waveguides up to arbitrary twist rates $(\bar{\kappa} r_c < r_c/ \rho < 0.13 \ll 1)$. However, at very high twist rates, the condition $2 r_c^x \ll 2 \pi \rho$ needs to be fulfilled as well since the ellipses would otherwise need to be curved as shown for an extreme twist rate of 100/mm in Fig.\ \ref{fgr:CrossSections}. The approximation therefore works best for small core radii. In this work we simulated waveguides up to twist rates of 20/mm where $2 r_c^x < 7.3 \ \upmu \mathrm{m} \ll \ 2 \pi \rho \approx 88.0 \ \upmu \mathrm{m}$, such that the approximation of the cross sections as ellipses is justified. \\

\subsection{Transformation of effective index to lab frame}

The results returned by the solver are the modal fields in the $xy$ plane at $z=0$ and the effective index $n_{\mathrm{eff}}^{\mathrm{Helical}}$ such that the fields $\vb{\tilde{F}}$ in the helicoidal coordinate system satisfy \cite{Weiss2013}:
\begin{equation}
	\vb{\tilde{F}(\xi_1,\xi_2,\xi_3)} = \mathrm{e}^{\mathrm{i} k \xi_3 \, n_{\mathrm{eff}}^{\mathrm{Helical}}} \vb{F(\xi_1,\xi_2)}.
\end{equation}
We apply a coordinate transformation to evaluate the fields in Cartesian coordinates at $z \neq 0$ which subsequently allows to display all modal quantities within the NB plane. Step-by-step instructions are available in Sec.\ S\rom{5} 
of the Supplemental Material \cite{SI_Ref}.

When transforming the fields to Cartesian coordinates, the fields develop an additional $z$-dependent phase factor which we refer to as the transformation phase (cf. Sec.\ S\rom{5} A 1 
ff. of the Supplemental Material \cite{SI_Ref}). We show that the transformation phase only increases linearly in $z$ under specific conditions. Most importantly, the following conditions have to be satisfied: (1) the electric field is circularly polarized with spin $s=\pm 1$, and (2) its spatial phase profile is flat or has an OAM profile with an $\mathrm{e}^{\mathrm{i} l \phi}$ phase dependence. In this case it is possible to define an effective index $n_{\mathrm{eff}}^{\mathrm{Lab}}$, such that the phase of the fields measured in the lab frame increases as $\mathrm{e}^{\mathrm{i} k z \, n_{\mathrm{eff}}^{\mathrm{Lab}}}$ with:
\begin{equation} \label{EffIndexTransformation}
	n_{\mathrm{eff}}^{\mathrm{Lab}} = n_{\mathrm{eff}}^{\mathrm{Helical}} + (s+l) \frac{\alpha \lambda}{2 \pi}.
\end{equation}
This equation holds both for off-axis twisted waveguides and on-axis twisted waveguides and matches with earlier derivations valid for on-axis twisted waveguides \cite{Weiss2013} and off-axis twisted waveguides \cite{Alexeyev2008} that were derived based on different approaches. Compared to these earlier derivations, we can also treat cases where the modes feature a noncircular polarization state and find that the transformation phase does not increase linearly in $z$ in this case. More intuitively, these phase changes are caused by the rotation of the polarization ellipse, following the twist of the waveguide. For example, if the short axis of the polarization ellipse points along the $x$ direction at $z=0$, it will point along the $y$ direction at $z=P/4$. This effect is unique to modes in twisted waveguides as the polarization state of an eigenmode in a straight waveguide does not change during propagation. The definition of an effective index in the lab frame for such elliptically polarized modes might therefore give the wrong impression, that they could couple to a mode in a straight waveguide with the same effective index. Keeping these caveats in mind, we still apply Eq.\ (\ref{EffIndexTransformation}) to all results for better comparability. \\

\subsection{Analytical description of the effective index of Frenet-Serret waveguides} \label{AnalyticalModelAndSO}

Helical waveguides of the Frenet-Serret type have been studied since the 1980s and a thorough theoretical model has been developed by Alexeyev and Yavorsky based on perturbation theory in 2008 \cite{Alexeyev2008}. The model is valid under four conditions: (1) The waveguide is weakly coiled ($r_c \bar{\kappa} \ll 1$), (2) the fields are transverse in the Frenet-Serret frame (scalar wave approximation), (3) the torsion $\tau$ of the helix can be treated as the small parameter in perturbation theory, and (4) the fields are circularly polarized and posses an OAM phase profile with topological charge $l$. In this case, the propagation constant of modes calculated within the Frenet-Serret coordinate frame is changed from that of the straight waveguide by $\Delta \beta^{\mathrm{Frenet}} = (s+l) \, \tau $ \cite{Alexeyev2008}. In the lab frame, the effective mode index can then be approximated as (see derivation in Sec.\ S\rom{6} 
of the Supplemental Material \cite{SI_Ref}):
\begin{equation} \label{EffIndexAlexeyev}
	n_{\mathrm{eff}}^{\mathrm{Lab}} = \underbrace{n_0 \sqrt{1+\alpha^2 \rho^2}}_{\substack{\text{Geometric increase} \\ \text{in path length}}} + \underbrace{(s+l) \frac{\alpha}{k_0} \left( 1- \frac{1}{\sqrt{1+\alpha^2 \rho^2}}\right)}_{\substack{\text{Spin-orbit $(s)$ and} \\ \text{orbit-orbit $(l)$ interaction}}}.
\end{equation}
For small twist rates $\alpha \rho \ll 1$ the result can be simplified showing that the splitting of modes with different total angular momentum increases strongly with twist rate and helix radius:
\begin{equation}
	n_{\mathrm{eff}}^{\mathrm{Lab}} = n_0 \sqrt{1+\alpha^2 \rho^2} + (s+l) \frac{\rho^2 \alpha^3 \lambda}{4 \pi} \text{\ \ \ for $\alpha \rho \ll 1$}.
\end{equation}
Their result shows that twisting a waveguide lifts the degeneracy between modes with the same magnitude of total angular momentum but different sign. In straight step-index fibers these modes are degenerate and therefore the spin and OAM state is not conserved during propagation (minor imperfections in any real-world waveguide lead to a coupling of the degenerate modes).
\begin{figure*}[t]
	\includegraphics[width=155mm ]{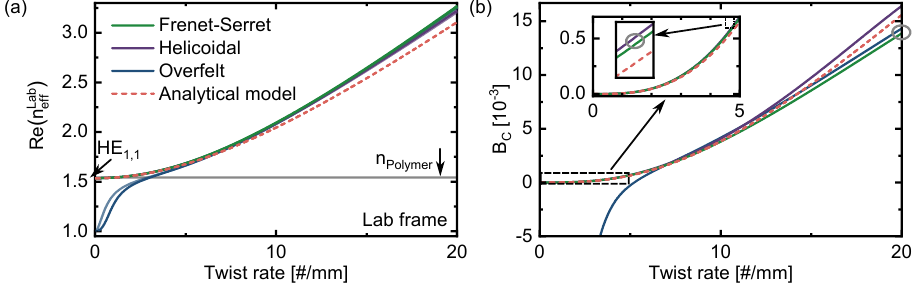}
	\caption{Optical properties of multimode helical waveguides ($r_c = 1.8 \ \upmu \mathrm{m}$). (a) Twist rate dependence of the real part of the effective mode index of the two fundamental modes calculated in the lab frame. A darker color shade denotes the LCP mode, a lighter shade the RCP mode. (b) Circular birefringence $B_C$ (i.e., difference between the effective index of the two lowest-order modes). Note, that for low twist rates (< 5/mm) the modes of the Overfelt waveguide become elliptically polarized. Red dashed lines in (a,b) represent an analytical prediction for the Frenet-Serret waveguide [Eq.\ (\ref{EffIndexAlexeyev})].} 
	\label{fgr:Multimodel0_Split0}
\end{figure*}

The birefringence caused by the helical path can also be derived on a more fundamental basis, using the "semi-geometrical optics" approximation \cite{Liberman1992, Bliokh2015}. Using this approximation, the motion of a wavepacket on scales much larger than the wavelength is influenced by spin-orbit and orbit-orbit interactions of light \cite{Fedoseyev2001, Bliokh2006a} (for more details see Sec.\ S\rom{6} A 
and Sec.\ S\rom{9} 
of the Supplemental Material \cite{SI_Ref}). These interactions result in the modes of helical waveguides acquiring a spin- and OAM-dependent Berry phase \cite{Chiao1986, Bliokh2015, Bliokh2006a}, given by the observation that light traces out a closed loop in $k$-space as it propagates through one turn of the helix. The solid angle $\Omega$ that this loop encloses determines the Berry phase as $\Phi_B = (s+l) \Omega$ and results in the same birefringence term stated in Eq.\ (\ref{EffIndexAlexeyev}) for the lab frame. Notably, the observation of circular birefringence in helically coiled fibers constitutes the first experimental confirmation of Michael Berry’s theoretically predicted geometrical phase factor \cite{Tomita1986, Chiao1986, Berry1984}.

\subsection{Simulation results for optical properties of helical waveguide geometries}
\subsubsection{Parameters of multimode waveguides}
With these preparations, we start analyzing the twist-rate-dependent optical properties of multimode variants of the three waveguides. Based on a recent realization of 3D-nanoprinted helical waveguides \cite{Gao2020} and related work on 3D-nanoprinted hollow-core waveguides \cite{Jain2019, Burger2021, Burger2022}, realistic parameters for this fabrication technique were chosen. In particular, $r_c = 1.8  \ \upmu \mathrm{m}$, $\lambda = 770 \ \mathrm{nm}$, and a refractive index of the core of $n_{\mathrm{co}} = 1.5423$ with a cladding made out of air ($V$ number: $17.25$, number of guided modes: $V^2/2 \approx 149$). These values correspond to a commonly used photoresist at this wavelength (IP-Dip, Nanoscribe GmbH).

\subsubsection{Fundamental modes in multimode waveguides} \label{MultimodeOptProperties}

The optical properties of the twisted multimode waveguides are very similar across a large range of investigated twist rates (5/mm - 20/mm) since the cross sections of the waveguides are so wide that the specific shape plays a minor role. The two fundamental modes are circularly polarized with the wavefronts lying in the NB plane as shown in Figs.\ S13 
and S14 
of the Supplemental Material \cite{SI_Ref}.

The real part of the effective index and the circular birefringence (difference in the real part of the effective index between the LCP and RCP mode, evaluated in the lab frame) match very well with the analytical model of Eq.\ (\ref{EffIndexAlexeyev}) for all waveguide types. Off-axis twisting has two immediate consequences: (1) the effective index of the mode increases due to the longer path that the light is traveling along the helix to reach a certain distance $z$ [first term in Eq.\ (\ref{EffIndexAlexeyev}) and Fig.\ \ref{fgr:Multimodel0_Split0}(a)], and (2) a splitting occurs between modes with different total angular momentum [second term in Eq.\ (\ref{EffIndexAlexeyev}) and Fig.\ \ref{fgr:Multimodel0_Split0}(b)]. Compared to typical values of birefringence found in polarization-maintaining fibers ($10^{-3}$-$10^{-4}$), much higher birefringence on the order of $10^{-2}$ can be reached at the highest investigated twist rate of 20/mm ($\alpha \rho = 1.8$).

Since the cross section of the Overfelt waveguide becomes infinitely narrow along the \mbox{B direction} for $\alpha \to 0$ (see Fig.\ S14 
of the Supplemental Material \cite{SI_Ref}), its optical properties deviate from those of the other two waveguides for twist rates below 5/mm. Due to the narrowing of the cross section, the two fundamental modes change from being circularly polarized to linearly polarized at low twist rates, resulting in a larger (linear) birefringence. As the modes are more and more localized in air, the effective index approaches 1 for $\alpha \to 0$. \\

\subsubsection{OAM modes in multimode waveguides}
\begin{figure*}[ht]
	\includegraphics[width=155mm]{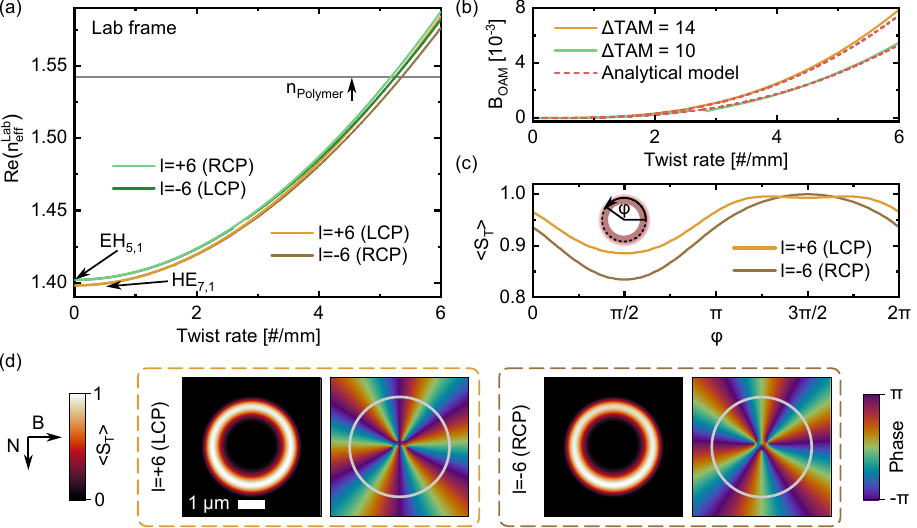}
	\caption{OAM modes in Frenet-Serret waveguide geometry ($r_c = 1.8 \ \upmu \mathrm{m}$). (a) Twist rate dependence of the real part of the effective mode index in the lab frame for the four modes with $\abs{l}=6$. The two modes with total angular momentum (TAM) of $\pm 5$ (green) stem from the even and odd $\mathrm{EH}_{5,1}$ modes of the untwisted waveguide, while the two modes with TAM of $\pm 7$ (orange) originate in the $\mathrm{HE}_{7,1}$ modes. (b) OAM birefringence between these mode pairs. Note, that the other pairings (+5/-7 and -7/+5) are already nondegenerate in the untwisted case. The simulated splitting matches well with the analytical prediction [red dashed line, Eq.\ (\ref{EffIndexAlexeyev})]. (c,d) Longitudinal component of the Poynting vector of the modes with TAM of $\pm 7$ for a twist rate of 5/mm. Evaluating the azimuthal distribution of the Poynting vector along a selected circle shows an increase in the intensity at the side of the waveguide that points towards the central twist axis. This asymmetry is different for the two modes, which might be a result of the photonic orbit-orbit interaction. Panel (d) additionally displays the phase of the B component of the electric field with the cross section of the waveguides shown as gray line.}
	\label{fgr:Multimodel6}
\end{figure*}
Next, we investigated the four OAM modes with $\abs{l}=6$ in the multimode Frenet-Serret waveguide. The modes are organized in two groups, one originating from the $\mathrm{EH}_{5,1}$ mode pair of the untwisted waveguide with a total angular momentum (TAM) of $\pm 5$ and the other one from the $\mathrm{HE}_{7,1}$ mode pair with a TAM of $\pm 7$ \cite{Ramachandran2013}. As the twist rate rises, the effective index of the modes overall increases with the same geometrical factor as the fundamental modes [first term in Eq.\ (\ref{EffIndexAlexeyev}) and Fig.\ \ref{fgr:Multimodel6}(a)]. On the other hand, the birefringence within each mode pair is 5 or 7 times larger than for the fundamental modes, respectively, as expected from the second term in Eq.\ (\ref{EffIndexAlexeyev}). This larger splitting for OAM modes is a consequence of the photonic orbit-orbit interaction term being $l$ times as large as the spin-orbit interaction term. Off-axis twisting therefore lifts the degeneracy of OAM modes with the same total angular momentum, resulting in OAM birefringence. This effect can be used to create OAM-maintaining fibers. Furthermore, twisting can induce coupling between modes, which enables twist-based mode converters \cite{Alexeyev2008}. While not explored further in this study, such a coupling can be observed for the mode pair with a TAM of $\pm 5$ at a twist rate of 2.7/mm, explaining the discontinuity in the green curves in Fig.\ \ref{fgr:Multimodel6}(a,b).

\subsubsection{Fundamental modes in single-mode waveguides} \label{ModesInSinglemodeWG}

\begin{figure}[h!]
	\includegraphics[width=77.5mm]{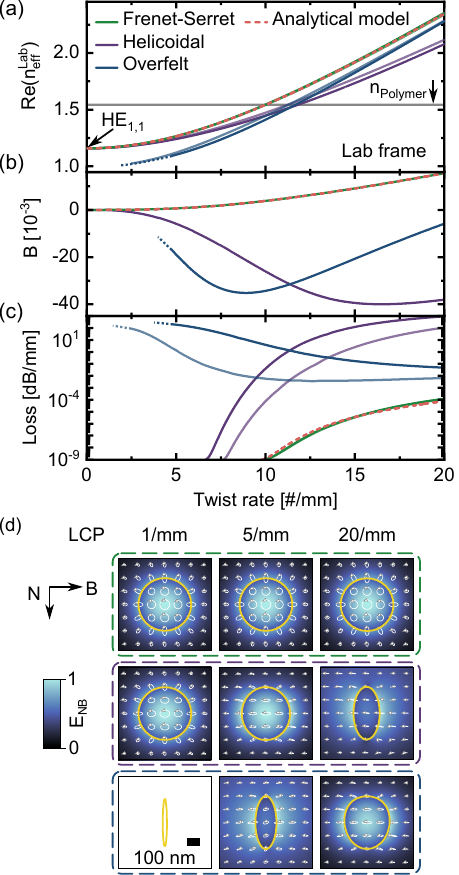}
	\caption{Optical properties of single-mode helical waveguides ($r_c = 0.2 \ \upmu \mathrm{m}$). (a,b,c) Twist rate dependence of the real part of the effective mode index, the birefringence, and the attenuation of the two fundamental modes calculated in the lab frame. A darker color shade denotes the LCP mode, a lighter shade the RCP mode. Red dashed lines in (a,b,c) represent an analytical prediction for the Frenet-Serret waveguide [Eqs.\ (\ref{EffIndexAlexeyev}),(\ref{eq:BendLoss})]. (d) Spatial distribution of the transverse (NB) component of the electric field of the LCP mode at three different twist rates. Polarization ellipses show that the field of the Frenet-Serret waveguide (green box) remains circularly polarized at all twist rates. Helicoidal waveguide (purple box) and Overfelt waveguide (blue box) feature elliptically polarized eigenmodes at high or low twist rate, respectively. The cross section of the waveguides is highlighted as yellow line. Note, that the modes of the Overfelt waveguide could not be calculated for twist rates below 5/mm due to very high loss and low confinement in the core (blue dots in a,b,c and blank field in d).}
	\label{fgr:Singlemodel0}
\end{figure}

To study single-mode variants of the three waveguide types, the core radius was reduced to $r_c = 0.2 \ \upmu \mathrm{m}$, resulting in a $V$ number of $V=1.92$. Contrary to the multimode case, the optical properties of these waveguides differ strongly between each other. As the cross section is much narrower, any change in the cross section from a circular profile affects the polarization of the mode as shown in Fig.\ \ref{fgr:Singlemodel0}(d). The resulting mix of linear and circular birefringence in the fundamental modes of the helicoidal and Overfelt waveguide can be much larger and of opposite sign than that of the Frenet-Serret waveguide [Fig.\ \ref{fgr:Singlemodel0}(b)]. As the cross section of the Frenet-Serret waveguide is circular in the NB plane at all twist rates, its modes remain circularly polarized and their effective indices are accurately described by Eq.\ (\ref{EffIndexAlexeyev}).
Another aspect to consider is that the fraction of power located in air is also dependent on the shape of the cross section. For example, the fraction of power in air is increased at small twist rates ($\alpha \rho \ll 1$) for the Overfelt waveguide, and at large twist rates ($\alpha \rho >1$) for the helicoidal waveguide, which results in a reduction of the effective index [Fig.\ \ref{fgr:Singlemodel0}(a) and Fig.\ S11 
of the Supplemental Material \cite{SI_Ref}]. As expected from the analytical formula for the cross sections of the waveguides (Table \ref{ParamsEllipsesMain}), the properties of the Frenet-Serret waveguide converge with those of the helicoidal waveguide for low twist rates and with those of the Overfelt waveguide for high twist rates. The propagation loss of the waveguides will be discussed below in Sec.\ \ref{SecLoss}.

\subsection{Effects of bending on modes in helical waveguides }
\subsubsection{Fundamental modes in multimode waveguides}

\begin{figure}[h]
	\includegraphics[width=77.5mm]{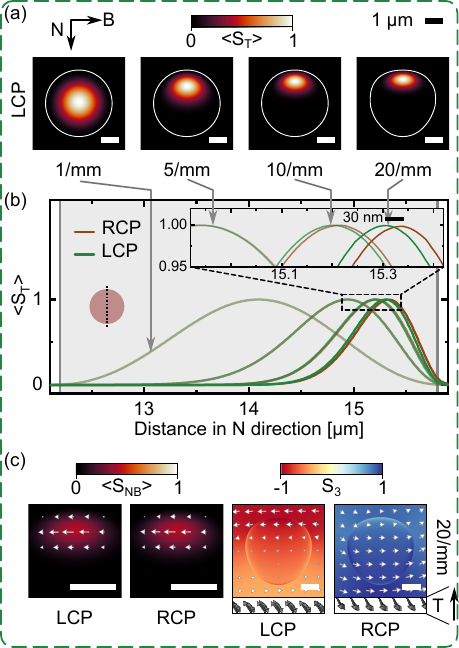}
	\caption{Spatial properties of modes in the multimode Frenet-Serret waveguide depicted in the TNB frame. (a) Distribution of the T component of the Poynting vector (i.e., tangential to the helix) of the LCP mode at four different twist rates. At increasing twist rate, the waveguide becomes bent such that the center of the mode moves to larger radii, i.e., away from the twist axis. A line cut of these distributions along the N direction through the center of the waveguide is shown in (b). The zoomed inset shows that the RCP mode (brown) moves further outwards than the LCP mode (green), potentially a consequence of the photonic spin Hall effect. (c) At a high twist rate of 20/mm, the Poynting vector develops a transverse component pointing in the negative B direction for both LCP and RCP modes. The spin vector of the electric field $\vb{s}_{\vb{E}}$ also develops a transverse component which points in opposite directions for the two polarizations. The magnitude of $\vb{s}_{\vb{E}}$ is equal to the third component of the Stokes vector $S_3$.} 
	\label{fgr:Multimodel0_Split1}
\end{figure}

Next, we studied the spatial properties of the modes in more detail with an overview of all modes available in Sec.\ S\rom{10} 
of the Supplemental Material \cite{SI_Ref}. Regarding the intensity distributions of the fundamental modes in the multimode waveguides, twisting has two effects: (1) the mode profile becomes narrower and (2) it shifts away from the twist axis. These effects are shown in Fig.\ \ref{fgr:Multimodel0_Split1}(b) for the Frenet-Serret waveguide and in Fig.\ S14 
for the helicoidal and Overfelt waveguides. Both effects are well known from fiber bends in a two-dimensional plane as the radius of curvature decreases \cite{Marcuse1976}. If the bent waveguide is approximated as a circle with a radius of curvature $R$, it can be mapped to a straight waveguide using a conformal transformation \cite{Heiblum1975}. This mapping results in a modified refractive index profile that increases approximately linear (for $R \gg r_c$) across the waveguide and cladding: $n \approx n_0 \, (1+x/R)$, where $x=0$ corresponds to the center of the waveguide and $n_0$ is the refractive index profile of the waveguide before bending \cite{Schermer2007}. More intuitively, as the mode has to propagate a larger distance on the outside of the bend than on the inside, the mapped refractive index increases away from the center of curvature. The reason for the observed shift of the modal patterns towards the region of higher index can be seen when noting that the scalar wave equation is equivalent to the time-independent Schrödinger equation \cite{Griffiths2018a} for a potential equal to $-n^2$. Consequently, the mode moves to larger radii to minimize its "energy". For helical waveguides the radius of curvature of the bend is given by $R= 1/\bar{\kappa}$, which decreases with increasing twist rate. \\

\subsubsection{OAM modes in multimode waveguides}

While twisting strongly impacts the effective index of OAM modes, its impact on the mode profile is much weaker than for the fundamental mode. Contrary to the fundamental modes, the center of the modes with $\abs{l}=6$ does not shift away from the twist axis as shown in Fig.\ \ref{fgr:Multimodel6}(d). This is to be expected since the effect of bending on the mode profile is known to decline as the mode order increases \cite{Marcuse1976, Jiao2019}. What is typically observed for OAM modes in bent fibers is that the angular intensity distribution is slightly nonuniform with a peak on the side facing away from the center of curvature \cite{Marcuse1976, Jiao2019, HaozheYan2017}. However, the OAM modes in the twisted Frenet-Serret waveguide peak on the side facing towards the twist axis as shown in Fig.\ \ref{fgr:Multimodel6}(c). This asymmetry must therefore be purely related to the twist. \\

\subsubsection{Fundamental modes in Single-mode waveguides}

The intensity distribution of the modes in the single-mode Frenet-Serret waveguide remains virtually unaffected by twisting due to the confinement provided by the high index contrast as shown in Fig.\ \ref{fgr:Singlemodel0_Appl}(a) and Fig.\ S16 
of the Supplemental Material \cite{SI_Ref}. For systems with a lower index contrast, a shift of the center of the mode would be expected \cite{Marcuse1976} similar to what has been discussed above for the multimode system. \\

\subsection{Loss in helical waveguides} \label{SecLoss}

Propagation loss in off-axis twisted waveguides is different from the loss of the corresponding untwisted waveguide for two reasons: (1) The geometric path length to reach a certain axial distance $z$ is increased because the light is traveling along a helical trajectory, and (2) additional loss arises from bending. The geometric factor effectively increases the intrinsic loss $\gamma_0$ of the untwisted waveguide, which might be present due to surface roughness or material absorption. We denote the associated attenuation coefficient for the twisted waveguide as $\gamma_{\mathrm{geo}}$  (the attenuation coefficient is generally defined via the relation $I(z) = I(0) \exp(-\gamma z)$, where $I$ denotes the optical power):
\begin{equation} \label{eq:GeoLoss}
	\gamma_{\mathrm{geo}} = \gamma_0 \, \frac{\sqrt{P^2+(2\pi \rho)^2}}{P}=  \gamma_0 \sqrt{1+\alpha^2 \rho^2}.
\end{equation}

Bend loss of optical fibers has been studied extensively for situations where the weak guidance approximation is satisfied ($\Delta \ll 1$), the modal field inside the twisted core is the same as in the straight fiber, and the radius of curvature is much larger than the radius of the core ($R \gg r_c)$ \cite{Snyder1983, Marcuse1976a}. More advanced models exist if one or more of these assumptions are not satisfied \cite{Marcuse1976, Marcatili1969, Smink2007}. Here, we use the model from Ref. \cite{Snyder1983} that is valid when all three assumptions are met and include a correction factor taking into account that the bends do not lie within a flat plane but occur along a helical path. In this case, the attenuation coefficient $\gamma_{\mathrm{bend}}$ is given by \cite{Soh2003}:
\begin{eqnarray} \label{eq:BendLoss}
	\gamma_{\mathrm{bend}} &=& \frac{\sqrt{\pi}}{2 r_c} \frac{V^2 \sqrt{W}}{U^2}  \sqrt{\frac{r_c}{R}} \ \mathrm{e}^{-\frac{4}{3} \frac{R}{r_c} \frac{W^3 \Delta}{V^2}} \nonumber\\ & &\times \underbrace{\left(1-\frac{1}{2} (\alpha \rho)^2+\frac{3}{32}(\alpha \rho)^4\right)}_{\mathrm{correction \ for \ helical \ bend}},
\end{eqnarray}
where $V=2 \pi (n_{\mathrm{co}}^2-n_{\mathrm{cl}}^2) r_c / \lambda $ is the $V$ number, $\Delta \  = (n_{\mathrm{co}}^2-n_{\mathrm{cl}}^2)/(2 n_{\mathrm{co}}^2) = 0.29$ is the refractive index contrast and $R=1/\bar{\kappa}$ is the radius of curvature of the helix. $U(V)$ and $W(V)$ are numerical solutions to a transcendental equation characterizing the mode and can be obtained from Ref. \cite{Snyder1983}. For the single-mode Frenet-Serret waveguide, one has $V=1.92$, $U=1.50$, and $W=1.20$ and for its multimode version $V=17.25$, $U=2.27$, and $W=17.10$. Note, that the mode profile of the multimode waveguide changes as the twist rate increases, and thus Eq.\ (\ref{eq:BendLoss}) is only approximately valid.

The resulting bend loss $\gamma_{\mathrm{bend}}$ is shown for the Frenet-Serret waveguide in Fig.\ \ref{fgr:Singlemodel0}(c), matching well with the simulated data. Interestingly, bending strongly affects the loss of the single-mode waveguide as the twist rate increases while it remains negligibly low for the multimode variant (below $2 \times 10^{-24} \mathrm{dB/mm}$ at the highest investigated twist rate). As the simulated loss of the multimode waveguides is below the noise level of the solver, it is not shown. This difference in bend loss can be understood based on the conformal transformation method described above. Since the mapped refractive index profile increases away from the center of curvature, at some distance from the waveguide - the caustic boundary - the mapped index of the cladding is higher than that of the core mode and the field becomes radiative due to the absence of total internal reflection \cite{Schermer2007}. As the field of the multimode waveguide remains well confined within the core, its amplitude at the caustic boundary is very low. In the single-mode waveguide on the other hand, a much larger fraction of the field is present inside the cladding, thus explaining the large difference in bend loss.

The geometric contribution to the loss $\gamma_{\mathrm{geo}}$ was not analyzed in this study because the material of the waveguide was assumed to be lossless ($\gamma_0=0$). Preliminary simulations with lossy materials showed that the total loss is well described as the sum $\gamma_{\mathrm{geo}}+ \gamma_{\mathrm{bend}}$ in this case.

\subsection{Twist-induced effects on spatial mode properties}

\subsubsection{Spin- and OAM-dependent effects}

\begin{figure}[h!]
	\includegraphics[width=77.5mm ]{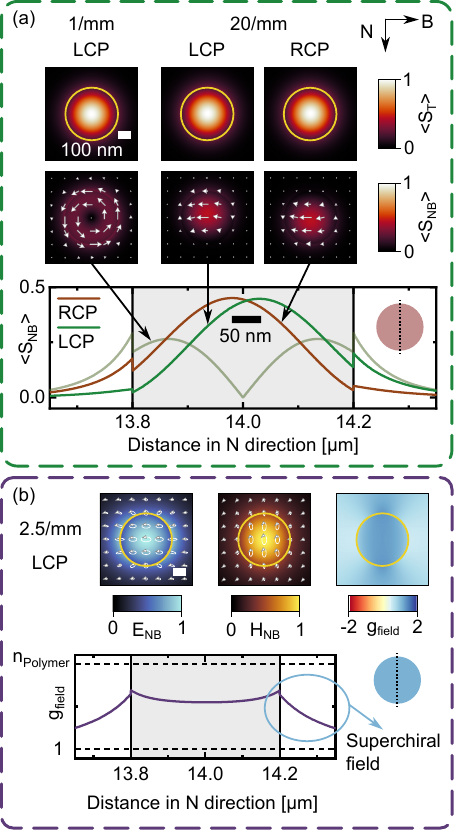}
	\caption{Spatial properties of modes in two single-mode helical waveguide geometries ($r_c = 0.2 \ \upmu \mathrm{m}$). (a) Distribution of the longitudinal (T) and transverse (NB) component of the Poynting vector at low (1/mm) and high (20/mm) twist rates for the Frenet-Serret waveguide. The longitudinal component remains unchanged when the twist rate increases while the transverse component shifts in opposite directions along the N axis. This splitting might be a consequence of the photonic spin Hall effect. (b) The elliptically polarized eigenmodes of the helicoidal waveguide feature a reduced electric field (left panel) on the top and bottom surface of the waveguide while the magnetic field is enhanced (middle panel). This difference results in a superchiral field with $g_\mathrm{field}$ being larger than the refractive index on the outside of the core (right panel). A line cut of this distribution along the N direction through the center of the waveguide is shown in the bottom panel. The yellow circles in (a,b) highlight the cross section of the waveguides.}
	\label{fgr:Singlemodel0_Appl}
\end{figure}

Apart from the large shift of the center of the modes induced by bending, we also observe several spin- and OAM-dependent splittings in the spatial properties of the modes, summarized here for the Frenet-Serret waveguide: (1) For the multimode waveguide, the center of the LCP and RCP modes (evaluated on the T component of the Poynting vector) are split along the N direction as shown in Fig.\ \ref{fgr:Multimodel0_Split1}(b). The splitting increases with twist rate and reaches 30 nm (about 1 \% of the core diameter) for a twist rate of 20/mm. (2) In the single-mode case, such a splitting occurs for the transverse component of the Poynting vector $\vb{S}_{\mathrm{NB}}$ while no splitting can be observed in its longitudinal component. The splitting reaches 50 nm (12.5 \% of the core diameter) for a twist rate of 20/mm as shown in Fig.\ \ref{fgr:Singlemodel0_Appl}(a). (3) For the OAM modes in the multimode variant, we observe that the difference between the intensity on the top and bottom side of the vortex depends on the sign of the total angular momentum. Bottom and top refers to the side facing towards and away from the twist axis, respectively as shown in Fig.\ \ref{fgr:Multimodel6}(c,d).

As a result, we hypothesize that these spin- and OAM-dependent splittings arise due to an interplay between the confinement provided by the waveguiding structure, and the photonic spin Hall \cite{Liberman1992, Bliokh2008} and photonic orbital Hall \cite{Fedoseyev2001, Bliokh2006a} effects. When light propagates along a curved trajectory which changes on lengthscales much larger than the wavelength, its movement can be characterized by equations of motion for the center of gravity of the mode. These equations contain a spin-orbit \cite{Liberman1992, Bliokh2015} and an orbit-orbit \cite{Bliokh2006a} interaction term, that results in spatial splittings between modes with distinct total angular momenta as described in more detail in Sec.\ S\rom{9} 
of the Supplemental Material \cite{SI_Ref}. In fact, these photonic spin Hall and orbital Hall effects are just another consequence of the spin-orbit and orbit-orbit interaction of light \cite{Bliokh2015}, which so accurately describes the circular birefringence of the helical waveguides (cf. Sec.\ \ref{AnalyticalModelAndSO}). Therefore it is likely that these effects are at the origin of the observed spatial splitting, although they do not apply directly to modes confined in waveguides. \\

\subsubsection{Tansverse components of the Poynting and spin vectors}

At high twist rates ($>$ 10/mm), the fundamental modes of all investigated waveguide types (single-mode and multimode) develop an increasingly large transverse component of the Poynting vector $\vb{S}_{\mathrm{NB}}$, shown exemplarily for the Frenet-Serret waveguide in Fig.\ \ref{fgr:Multimodel0_Split1}(c) and Fig.\ \ref{fgr:Singlemodel0_Appl}(a). The direction of this additional component is along the negative B direction regardless of polarization. Similarly, the spin vector $\vb{s}_{\vb{E}}$ of the electric field contains a transverse component along the B direction for high twist rates, pointing in opposite directions for the two spin states. The evolution of these properties at different twist rates can be found for all waveguide types in Sec.\ S\rom{10} 
of the Supplemental Material \cite{SI_Ref}. Transverse spin is well known to occur in evanescent waves where its direction is independent of polarization and led to applications involving spin-momentum locking \cite{Bliokh2014, Aiello2015, Lodahl2017}. The origin of transverse spin and momentum in helical waveguides remains to be studied. \\

\subsubsection{Elliptically polarized eigenmodes and superchiral fields in helicoidal waveguides}

Lastly, we want to point out an effect that becomes apparent when the cross section of the twisted single-mode waveguide is elliptical in the NB plane as it is the case for the helicoidal waveguide. As discussed, the interplay of the linear birefringence of the core and the circular birefringence caused by the twist results in elliptically polarized eigenmodes. Similar to a linearly polarized mode, the modal fields are enhanced in the direction of the long axis of the polarization ellipse. These locations differ for the electric and magnetic field as shown in Fig.\ \ref{fgr:Singlemodel0_Appl}(b). In this example, the magnetic field is enhanced at the top and bottom surface of the core, while the electric field is reduced. Combined with the fact that the fields are still circularly polarized to a sufficient degree, such a configuration is ideal for creating superchiral fields. We quantify the superchirality in terms of the factor $g_\mathrm{field}$ (defined and explained in Sec.\ S\rom{8} 
of the Supplemental Material \cite{SI_Ref}), with $\abs{g_\mathrm{field}}>n$ indicating a superchiral field. In brief, $g_\mathrm{field}$ is the enhancement factor of the molecular $g$-factor, which is typically measured in chiral sensing experiments. Here, we found an enhancement factor of $\approx 1.4$ on the surface of the waveguide. Values of $g_\mathrm{field}$ for other twist rates can be found in Fig.\ S12 
of the Supplemental Material \cite{SI_Ref}. We note that this value is lower than in the pioneering experiment for superchiral fields, where an enhancement of around 11 was measured \cite{Tang2010, Tang2011}. However, our enhancement occurs in a region of space where the field has an appreciable intensity while the pioneering experiment was carried out in the node of a standing wave where the intensity is low. Further research is required to optimize the superchirality in helical waveguides (e.g., by increasing both the twist rate and the linear birefringence), and to find a method to place molecules precisely in the regions of superchiral field.

\section{Discussion}

The numerical results for the effective index of all three multimode helical waveguides are in high agreement with the analytical prediction, underpinning the fact that the cause of the circular birefringence and OAM birefringence is a purely geometrical effect inherent to the helical path. As such, the phase difference between modes with different total angular momentum $\delta n_{\mathrm{eff}}^{\mathrm{Lab}} k_0 z$ does not depend on wavelength, material, and core size [cf. Eq.\ (\ref{EffIndexAlexeyev})]. Helical waveguides can therefore find applications as broadband spin- and OAM-preserving waveguides. More generally, any waveguide with nonzero torsion at each point of its trajectory will be able to preserve the angular momentum state of the light. 

A typical example of a Frenet-Serret waveguide is a piece of fiber helically coiled around a cylinder, as it was used in the first experiments on helical waveguides \cite{Qian1986, Ross1984, Smith1978a, Tomita1986}. Helical waveguides can also be created by twisting a fiber with an off-axis core \cite{Varnham1985,Ma2011a}. Such fibers are fabricated by either spinning the preform while drawing the fiber or in a thermal post-processing step \cite{Russell2017}. In this situation, it is conceivable that both helicoidal or Frenet-Serret type waveguides can in principle be created depending on the fabrication conditions. Since both types converge to the same shape at low twist rates, a difference would only be observable at high twist rates ($\alpha \rho \gg 1$), which have not yet been experimentally realized (see Table S\rom{1} of the Supplemental Material \cite{SI_Ref}). Finally, the Overfelt waveguide might be relevant for situations where a helical waveguide is constructed by extending a toroidal shape in the $z$ direction or to describe Frenet-Serret waveguides at large twist rates.

All waveguides created with planar fabrication techniques on the other hand, necessarily have zero torsion and can therefore not exhibit circular or OAM birefringence based on geometric effects. To realize complex 3D waveguides directly on-chip, 3D-nanoprinting via two-photon-polymerization has emerged over the last decade as a promising alternative to planar fabrication techniques \cite{Lindenmann2012a, Schumann2014, Gao2020}. This technique can be used to realize all three helical waveguide types discussed in this work and provides the ability to freely choose the cross section within the NB plane. While the twist rate achievable in fibers is inherently limited by the properties of the glass, 3D-nanoprinting can potentially realize higher twist rates with pitch distances down to the core diameter. Here, we explored waveguides with pitch distances as low as 14 times the core diameter, but even lower pitch distances seem theoretically feasible given the low bend loss in the multimode versions of the waveguides. Preliminary experiments showed that twist rates of 10/mm can be reached experimentally, corresponding to $\alpha \rho = 0.9$ while a value of $\alpha \rho \approx 0.5$ was reported for a different design in \cite{Gao2020}. Another technique that allows realizing helical waveguides on a chip is direct laser writing, where a small refractive index modification is created in glass by scanning of a focused femtosecond laser beam \cite{Rechtsman2013, Szameit2010, Gross2015}. However, precise control over the cross section of the waveguide proved to be challenging due to aberrations of the focal spot inside the glass that need to be compensated \cite{Salter2012, Salter2019}.

The study of the polarization properties of modes in helical waveguides revealed that the shape of the cross section in the NB plane strongly impacts their optical properties if the waveguides are single-mode. Any deviation from a circular cross section can induce linear birefringence, thus creating elliptically polarized eigenmodes. The simple analytical formula by Alexeyev and Yavorsky \cite{Alexeyev2008} is therefore not sufficient to describe the effective index in this specific case. Instead, a recently developed theoretical model for off-axis twisted waveguides with elliptical cross sections in the Frenet-Serret frame \cite{Chen2021} could be applied to predict the birefringence and polarization of the helicoidal and Overfelt waveguide. This would require prior knowledge of the modes of an untwisted waveguide with the same elliptical cross section. Additionally, a recent perturbative model studied the evolution of polarization in a helical waveguide when excited by the fundamental mode of a straight waveguide, which corresponds to the typical situation in experiments \cite{Mieling2023}.

Elliptically polarized eigenmodes also occur in on-axis twisted waveguides with an elliptical cross section and were applied in creating ultranarrow (sub-megahertz) spectral dips in stimulated Brillouin scattering \cite{Choksi2022}. The authors used that the polarization state of elliptically polarized modes depends on the wavelength, which is another intriguing effect that is unique to twisted waveguides.

\section{Conclusion}

Using comprehensive theoretical and numerical analysis, we revealed the differences between three common helical waveguide geometries, which naturally result from the use of the Frenet-Serret, helicoidal, or Overfelt frame. While the Frenet-Serret waveguide maintains a circular cross section in the NB plane at all twist rates, the helicoidal and Overfelt waveguides can exhibit elliptical cross sections, leading to non-circular polarization if the core size is small. Such elliptically polarized modes were found to generate superchiral fields on the surface of the waveguides. Conversely, the effective index of all circularly polarized waveguide modes can be accurately described by an analytical formula by Alexeyev and Yavorsky \cite{Alexeyev2008} [Eq.\ (\ref{EffIndexAlexeyev})] even if (1) the index contrast is large ($n=1.54$ to air, $\Delta = 0.29$), (2) the mode is not entirely transversely polarized, (3) the mode profile is different from that of the untwisted waveguide, and (4) the twist rate approaches relatively high values. Propagation loss was found to consist of bend loss \cite{Soh2003} and intrinsic loss.

Additionally, we derived a transformation of the effective index from the helicoidal to the lab frame applicable to both on- and off-axis twisted waveguides up to arbitrary twist rates. We note that such a transformation should not be performed if the modes are elliptically polarized. Lastly, the study explored spin- and OAM-dependent splittings in the spatial intensity distribution of the modes, suggesting links to the photonic spin Hall and orbital Hall effects.

Traditionally, helical waveguides were realized by twisting glass fibers, limiting the achievable twist rates and lacking precise control over waveguide cross section. However, emerging fabrication techniques such as 3D-nanoprinting \cite{Gao2020, Jain2019, Burger2022} or direct laser writing hold promise for overcoming these limitations and enabling novel chip-integrated applications of helical waveguides, including chiral spectroscopy, integrated Brillouin lasing for ultrahigh-resolution optical sensing \cite{Choksi2022}, and OAM-maintaining transport of optical signals. On a fundamental level, exploring ultrahigh twist rates with pitch distances close to the diameter of the off-axis core may offer insights into complex physical effects such as strong spin-orbit and orbit-orbit interactions.

\appendix*
\section{Numerical simulations}

All waveguides were simulated in the helicoidal coordinate system using a commercial FEM solver (PropagatingMode module of JCMwave). The waveguide core is surrounded by air, followed by a perfectly matched layer (PML) to absorb outgoing power resulting from bending of the waveguide. The helicoidal coordinate system is natively supported by the solver, including an appropriate definition of the PML. Convergence of the solver in terms of mesh size in core and air regions, and the distance between the waveguide and PML has been checked.

\begin{acknowledgements}
	We thank Adrian Lorenz (IPHT Jena) for helpful discussion on bending effects in fibers. Furthermore, we thank Sven Burger and Martin Hammerschmidt (JCMwave GmbH) for quickly patching an error in JCMwave that affected the calculation of fields at high twist rates.
	
	The authors acknowledge financial support from the German Research Foundation via the Grants MA 4699/2-1, MA 4699/9-1, SCHM2655/11-1, SCHM2655/15-1, SCHM2655/8-1, SCHM2655/22-1, WE 5815/5-1. S.A.M. additionally acknowledges the Lee-Lucas Chair in Physics.
\end{acknowledgements}

\bibliography{1_Bib}

\end{document}



\title{Supplemental material: \\ Impact of coordinate frames on mode formation in twisted waveguides}




\author{Johannes Bürger}
\email{j.buerger@physik.uni-muenchen.de}
\affiliation{Chair in Hybrid Nanosystems, Nanoinstitute Munich, Ludwig-Maximilians-Universität Munich, Königinstraße 10, 80539 Munich, Germany}

\author{Adrià Canós Valero}
\affiliation{Institute of Physics, University of Graz, Universitätsplatz 5, 8010 Graz, Austria}

\author{Thomas Weiss}
\affiliation{Institute of Physics, University of Graz, Universitätsplatz 5, 8010 Graz, Austria}
\affiliation{4th Physics Institute and SCoPE, University of Stuttgart, Pfaffenwaldring 57, 70569 Stuttgart, Germany}

\author{Stefan A. Maier}
\affiliation{School of Physics and Astronomy, Monash University, Clayton, Victoria 3800, Australia}
\affiliation{Chair in Hybrid Nanosystems, Nanoinstitute Munich, Ludwig-Maximilians-Universität Munich, Königinstraße 10, 80539 Munich, Germany}
\affiliation{The Blackett Laboratory, Department of Physics, Imperial College London, London SW7 2AZ, United Kingdom}

\author{Markus A. Schmidt}
\affiliation{Leibniz Institute of Photonic Technology, Albert-Einstein-Str. 9, 07745 Jena, Germany}
\affiliation{Otto Schott Institute of Materials Research (OSIM), Friedrich-Schiller-Universität Jena, Fraunhoferstr. 6, 07743 Jena, Germany}
\affiliation{Abbe Center of Photonics and Faculty of Physics, Friedrich-Schiller-Universität Jena, Max-Wien-Platz 1, 07743 Jena, Germany}

\date{April 5, 2024}

\maketitle


Here, we report additional results and relevant background information concerning an overview of related works on off-axis twisted waveguides (Sec.\ \ref{SI_PaperOverview}), the definition of the used optical parameters (Sec.\ \ref{SI_DefsOptQuantities}), the coordinate systems (Sec.\ \ref{SI_CoordSystems}), the approximation of the cross sections of the waveguides as ellipses (Sec.\ \ref{SI_CrossSecFormula}), properties of the coordinate transformation between the helicoidal coordinate system and the Cartesian laboratory frame (Sec.\ \ref{SI_TransformationGeneral}), an analytical description of the effective index of a Frenet-Serret waveguide by Alexeyev and Yavorsky (Sec.\ \ref{SI_Alexeyev}), the twist-rate dependence of the fraction of power present inside the core of single-mode helical waveguides (Sec.\ \ref{SI_FracPowerCore}), superchiral fields (Sec.\ \ref{SI_Superchiral}), the photonic spin Hall and orbital Hall effects (Sec.\ \ref{SI_SHE}), and detailed plots of the spatial properties of all investigated waveguide modes (Sec.\ \ref{SI_AllPlots}). 

\section{Comparison of works studying twisted waveguides with off-axis elements} \label{SI_PaperOverview}

An overview of works on off-axis twisted waveguides is provided in Table \ref{PaperOverviewTable_SI} on the next page for a comparison of the maximally achieved twist rates. Note, that off-axis twisted rods or capillaries can also occur in the cladding of on-axis twisted waveguides like twisted photonic crystal fibers (PCF). The analysis provided in this study could therefore have implications for all the listed works. \newpage

\begin{table} 
	\caption{Comparison of works on twisted waveguides.} \label{PaperOverviewTable_SI}
	\begin{ThreePartTable}
		\begin{ruledtabular}
		\begin{tabular}{p{0.30\linewidth}p{0.25\linewidth}p{0.06\linewidth}p{0.07\linewidth}p{0.1\linewidth}p{0.04\linewidth}p{0.15\linewidth}}
			Type                                       & Fabrication method     & Twist period $P \, [\upmu \mathrm{m}]$ & Radius $\rho \, [\upmu \mathrm{m}]$ & $\alpha \rho$      &Year & Ref. \\ \hline
			\multicolumn{7}{@{}l}{\textbf{Theory  \& simulation}}  \\[0.75mm] 
			Frenet-Serret, helicoidal, and Overfelt waveguide                                   & -                                     & 50                               & 14                         & 1.76             & 2023          & \textbf{This work}     \\ 
			Frenet-Serret waveguide        & -                                     & 118\tnote{c}                            & 3.5\tnote{c}                   & 0.19             & 2021          & Chen \cite{Chen2021}      \\ 
			Twisted solid-core PCF                                                             & -                                     & 622                               & 12\tnote{a}                    & 0.12       & 2013          &  Weiss \cite{Weiss2013}     \\[2mm] 
			\multicolumn{7}{@{}l}{\textbf{On-axis twisted waveguides with off-axis rods}}     \\[0.75mm]
			Solid-core PCF                 & Post-processing (CO2 laser)   & 341                              & 16\tnote{a}               & 0.29       & 2012          & Wong \cite{Wong2012}      \\ 
			Solid-core PCF                     & Post-processing (CO2 laser)   & 435                              & 16\tnote{a}                   & 0.23       & 2013          & Xi \cite{Xi2013a}       \\ 
			Solid-core PCF             & Post-processing               & 456                              & 16\tnote{a}                   & 0.22       & 2015          & Wong \cite{Wong2015}      \\ 
			Solid-core PCF         & Post-processing (CO2 laser)   & 622                              & 16\tnote{a}                   & 0.16       & 2013          & Xi \cite{Xi2013}       \\ 
			Solid-core PCF (3 cores) & Post-processing (CO2 laser)   & 5000                             & 3\tnote{b}$\, / $16\tnote{a}                 & 0.004/0.02 & 2014          & Xi \cite{Xi2014}        \\ 
			Coreless PCF                          & Preform spinning                      & 2000                             & 41\tnote{a}                   &                  & 2016          & Beravat \cite{Beravat2016}   \\ 
			Solid-core PCF (6 cores)   & Preform spinning                      & 2200                             & 6\tnote{b}$\, / $18\tnote{a}                       & 0.02/0.05        & 2017          & Russell\cite{Russell2017}   \\ 
			Solid-core PCF (3 cores)                                            & Preform spinning                      & 2500                             & 2.5\tnote{a}                        & 0.006            & 2020          & Loranger \cite{Loranger2020}  \\ 
			Coreless PCF                            & Preform spinning                      & 3600                             & 41\tnote{a}                    & 0.07       & 2019          & Roth \cite{Roth2019}      \\ 
			Solid-core PCF (3 cores) & Preform spinning                      & 5000                             & 5.2\tnote{b}$\, / $16\tnote{a}               & 0.007/0.02 & 2022          & Zeng \cite{Zeng2022}      \\ 
			Hollow-core PCF (single ring)                            & Preform spinning                      & 11900                            & 48\tnote{a}                         & 0.025      & 2018          & Roth \cite{Roth2018}      \\ 
			Hollow-core PCF (single ring)                         & Preform spinning                      & 12400                            & 34\tnote{a}                         & 0.017      & 2017          & Edavalath \cite{Edavalath2017} \\[2mm] 
			\multicolumn{7}{@{}l}{\textbf{Off-axis twisted waveguides}}   \\[0.75mm] 
			Off-axis twisted waveguide                                                         & 3D-nanoprinting                       & 500                        & 40                   & 0.5        & 2020          & Gao \cite{Gao2020}       \\
			Chiral long period grating                                                  & Post-processing (oven)        & 589                              & 52                         & 0.55             & 2010          & Kopp \cite{Kopp2010}      \\
			Helical core fiber                                                                 & Preform spinning                      & 2000                             & 184                        & 0.58             & 1985          & Varnham \cite{Varnham1985}   \\ 
			Chirally-coupled-core fiber                    & Preform spinning                      & 6100                             & 27                         & 0.03             & 2011          & Ma \cite{Ma2011a}       \\ 
			Helical waveguide array                                                            & Direct laser writing in glass         & 10000                            & 10                         & 0.006            & 2018          & Stutzer \cite{Stutzer2018}   \\
			Fiber wound around cylinder                                            & Macroscopic winding                                     & 12000                      & 48000                & 25         & 1984          & Ross \cite{Ross1984}      \\ 
			Fiber wound around cylinder                                            & Macroscopic winding                                     & 300000                           & 280000                     & 5.9              & 1986          & Tomita \cite{Tomita1986}      \\
		\end{tabular}
		\begin{tablenotes}
			\raggedright
			\item[a] Radius corresponding to outer air hole/capillary.
			\item[b] Radius corresponding to off-axis core.
			\item[c] For a wavelength of 770 nm. 
		\end{tablenotes}
	\end{ruledtabular}
	\end{ThreePartTable}
\end{table}

\section{Definitions of optical quantities} \label{SI_DefsOptQuantities}
Since different conventions exist in the literature for the definition of several optical quantities, the definitions used throughout this work are listed here.
\subsection{Spin and orbital angular momentum}
We use the following conventions for spin and orbital angular momentum. A forward propagating circularly polarized plane wave with spin $s = \pm 1$ is defined as:
\begin{align} \label{CirPol}
	\vb{E} = E_0 \Re{\ket{s} \mathrm{e}^{\mathrm{i} (k z -\omega t)}}, && 
	\ket{s} = 	\frac{1}{\sqrt{2}} \begin{pmatrix}
		1 \\
		s i  \\
		0 
	\end{pmatrix}, && \parbox{7em}{$s=+1$ for LCP \\ $s=-1$ for RCP}.
\end{align}
This definition is in line with the convention typically used in optics which defines the sense of rotation of the field vector as seen from the point of view of the receiver (looking into the beam) \cite{Feynman2011}. Right circular polarization (RCP) is then defined as a clockwise rotation of the polarization vector in time and left circular polarization (LCP) as an anticlockwise rotation.

Furthermore, we use the intensity normalized Stokes parameter $\hat{S}_3$, which is defined as \cite{Born2019}:
\begin{equation}
	\hat{S}_3 = \frac{I_{\mathrm{RCP}}-I_{\mathrm{LCP}}}{I} \ \ \xrightarrow{p=1} \ \  \parbox{8em}{$\hat{S}_3 = -1$ for LCP \\ $\hat{S}_3 = +1$ for RCP},
\end{equation}
where $I_{\mathrm{RCP}}$ and $I_{\mathrm{LCP}}$ are the intensities measured after a circular polarizer set to transmit RCP or LCP, respectively. $I$ is the total intensity and $p=1$ refers to fully polarized beams. Note, that $\hat{S}_3$ and spin have opposite signs with this commonly used definition.

For the spin vector $\vb{s}$ of the electric and magnetic field, respectively, we use the following definition \cite{Bliokh2014, Gil2021}:
\begin{equation}
	\vb{s}_{\vb{E}} = \Im{\vb{\hat{E}^*} \cross \vb{\hat{E}} }, \hspace{23 mm} \vb{s}_{\vb{H}} = \Im{\vb{\hat{H}^*} \cross \vb{\hat{H}} },
\end{equation}
with the normalized field vectors (i.e., of unit length) $\vb{\hat{E}}$ and $\vb{\hat{H}}$. Since the spin vector is oriented perpendicular to the polarization ellipse, it contains additional information about the polarization structure of the mode. The degree of circularity of the respective field is then given by:
\begin{equation} \label{StokesS3Mag}
	\abs{\hat{S}_3} =  \abs{\vb{s}}.
\end{equation}
The direction of the spin vector relative to the propagation direction of the mode is used to set the correct sign for $\hat{S}_3$. If the spin vector contains a component parallel to the propagation direction a negative sign is used for $\hat{S}_3$, if it contains an antiparallel component a positive sign is used. \\

The sign of the orbital angular momentum (OAM) is specified based on the phase distribution of the fields. A phase distribution proportional to (or resembling) an $\mathrm{e}^{\mathrm{i} l \phi}$ phase factor, carries an OAM of $l$ \cite{Allen1992}. This definition is in agreement with the definition of spin $s$ used above, such that the total angular momentum can be obtained as $s+l$ \cite{Marrucci2006}.

\subsection{Poynting Vector}

The time-averaged Poynting vector $\langle \vb{S} \rangle$ is calculated from the obtained field solutions as \cite{Born2019}:
\begin{equation}
	\langle \vb{S} \rangle = \frac{1}{2} \Re{\vb{E} \cross \vb{H}^*}.
\end{equation}

\section{Coordinate Systems for Twisted Waveguides} \label{SI_CoordSystems}

A helical waveguide can best be analyzed in a coordinate system that follows the twist of the structure such that the waveguide becomes invariant along one of the new coordinates. Here, three different choices of a suitable coordinate system are introduced: the Frenet-Serret frame, the helicoidal coordinate system, and a coordinate system used by Overfelt \cite{Overfelt2001}. All simulations for this work were performed using the helicoidal coordinate system.

\subsection{Frenet-Serret Frame} \label{FrenetSerretSec}

Any differentiable curve $\vb{c}$ that is parametrized by a single parameter $t$ can be described in a 'self-defined' local orthonormal coordinate system, the Frenet-Serret frame. \\

In three dimensions, the Frenet-Serret frame consists of the tangent vector $\vb{\hat{T}}$ at a given point of the curve, the normal vector $\vb{\hat{N}}$, and the binormal vector $\vb{\hat{B}}$ which are defined as follows \cite{Kuhnel2015}:
\begin{align} \label{FrenetCoord}
	\vb{\hat{T}} = \frac{\vb{c}'(t)}{\norm{\vb{c}'(t)}}, &&
	\vb{\hat{N}} = \frac{\vb{\hat{T}}'(t)}{\norm{\vb{\hat{T}}'(t)}}, &&
	\vb{\hat{B}} =  \vb{\hat{T}} \times \vb{\hat{N}}.
\end{align}
The derivatives of the basis vectors are governed by the Frenet equations \cite{Kuhnel2015}:
\begin{align} \label{FrenetEq}
	\underbrace{\vb{\hat{T}}'(t)  = C \bar{\kappa}  \vb{\hat{N}}}_{(a)}, &&
	\underbrace{\vb{\hat{N}}'(t)  = C(-\bar{\kappa}  \vb{\hat{T}} + \tau \vb{\hat{B}})}_{(b)}, &&
	\underbrace{\vb{\hat{B}}'(t)  =   - C \tau \vb{\hat{N}}}_{(c)},
\end{align}
where $C = \norm{\vb{c}'(t)}$. Since the definition of the Frenet-Serret frame is independent of the parametrization, $C$ can be chosen to be 1 if the curve is parametrized by the arc length. Eqs.\ (\ref{FrenetEq}a) and (\ref{FrenetEq}c) serve as the definition of two proportionality constants, the curvature $\bar{\kappa}$ \footnote{usually curvature is denoted by a plain $\kappa$, the bar was added to avoid confusion with the helical propagation constant $\kappa$ defined later} and the torsion $\tau$. Curvature $\bar{\kappa} (t)$ describes the deviation of the curve from a straight line, which can locally be described as a circle with radius \footnote{$1/\bar{\kappa}$ is the radius of curvature} $1/\bar{\kappa}$ where $\vb{\hat{N}}$ points towards its center. Torsion $\tau (t)$ is a measure of twist - in the sense that it describes how fast the binormal vector $\vb{\hat{B}}$ rotates around the tangent $\vb{\hat{T}}$. For a curve defined on a 2D plane, both $\vb{\hat{T}}$ and $\vb{\hat{N}}$ lie within this plane meaning that $\vb{\hat{B}}$ is perpendicular to this plane and never changes direction. Therefore, a curve having zero torsion at every point is equivalent to the curve lying in a 2D plane.

\subsubsection{Helices} \label{HelixParamSec}

A helix is a special type of curve which is defined by a constant curvature $\bar{\kappa}$ and torsion $\tau$ at every point of the curve \cite{Kuhnel2015}. The main parameters of a helix, which are used throughout this work are listed here in Table \ref{HelixDef}.
\begin{table}[H]
	\caption{\label{HelixDef} Definition of parameters used to describe a helix. Any curve with constant curvature and torsion is a helix \cite{Kuhnel2015}. Based on the definitions in Eq.\ (\ref{FrenetEq}c), $\tau$ is positive for a right-handed helix and negative for a left-handed helix. The sign of $\alpha$ depends on the choice of the coordinate system and will be defined below in each subsection.}
	\centering
	\begin{tabular}{rlrlrl}
		\hline\hline
		\rule{0pt}{4ex}
		Pitch: & $P$ \ \ \ \  & Twist rate: & $1/P$ & Curvature:  & $\bar{\kappa} = \frac{\rho}{\rho^2+\left(\frac{P}{2 \pi}\right)^2}$    \\[5mm]
		Radius:   & $\rho$ & Angular twist rate:   & $\abs{\alpha}= 2 \pi/P$  \ \ \ \       & Torsion: & $\abs{\tau} = \frac{2 \pi P}{\left(2 \pi \rho\right)^2+P^2}$ \\[3mm]
		\hline\hline
	\end{tabular}

\end{table}

All simulations in this work are performed for left-handed helices. A helix is left-handed when it is possible to move the left thumb along the helix axis such that the remaining fingers curl along the helix trajectory. This definition is in line with the definition of circular polarization in Eq.\ (\ref{CirPol}): the curve traced out by the tip of the polarization vector of an LCP beam in space (at a fixed moment in time) is a left-handed helix.
\begin{figure}[h!]
	\centering
	\includegraphics[scale=1]{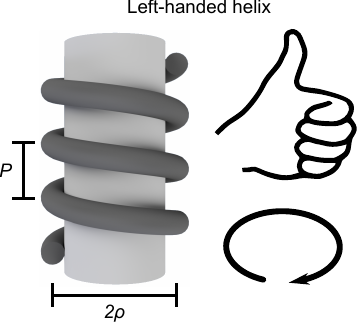}
	\caption[]{Definition of a left-handed helix with pitch $P$ and radius $\rho$. When moving the thumb of the left hand along the helix axis and rotating the hand in the direction of the curled fingers, the fingertips move along the path of a left-handed helix. Note, that the same result is obtained when pointing the thumb downwards. Drawing of hand was adapted from \cite{LeftHand}.}	
\end{figure}
\newpage
The unit vectors of an off-axis twisted waveguide following a left-handed helical path $\vb{c}(t)$ with $\alpha >0$ can be calculated according to Eq.\ (\ref{FrenetCoord}) yielding:
\begin{equation}
	\vb{c}(t) 
	=
	\begin{pmatrix}
		\rho \cos(\alpha t) \\
		-\rho \sin(\alpha t) \\
		t
	\end{pmatrix},
\end{equation}
\begin{align}
	\vb{\hat{T}}
	=
	\frac{1}{\sqrt{1+(\alpha \rho )^2}}
	\begin{pmatrix}
		-\alpha \rho  \sin(\alpha t) \\
		-\alpha \rho  \cos(\alpha t) \\
		1
	\end{pmatrix}, &&
	\vb{\hat{N}} 	=
	\begin{pmatrix}
		-\cos(\alpha t) \\
		\sin(\alpha t) \\
		0
	\end{pmatrix}, &&
	\vb{\hat{B}} 	=
	\frac{1}{\sqrt{1+(\alpha \rho )^2}}
	\begin{pmatrix}
		-  \sin(\alpha t) \\
		-  \cos(\alpha t) \\
		-\alpha \rho
	\end{pmatrix}.
\end{align}
These orthonormal unit vectors are shown below. $\vb{\hat{T}}$ points in the tangential direction, $\vb{\hat{N}}$ lies in the $xy$ plane and points towards the twist axis, and $\vb{\hat{B}}$ is orthogonal to both. Therefore, a helical waveguide can be created by defining a refractive index profile in the NB plane and extending it infinitely along the T direction. The simplest helical waveguide consists of a circular cross section in the NB plane and is referred to as a "Frenet-Serret waveguide" in this work.
\begin{figure}[h!]
	\centering
	\includegraphics[scale=1]{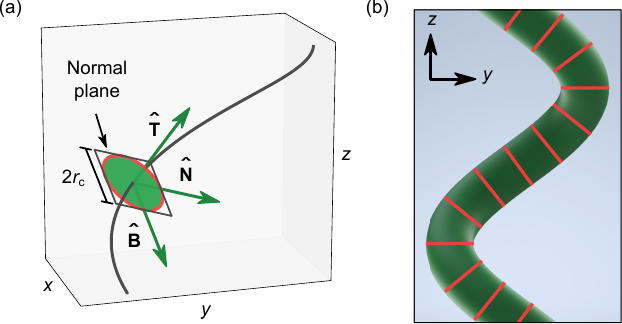}
	\caption[]{Definition of a helical waveguide in the Frenet-Serret frame. The basis vectors of the Frenet-Serret frame for a left-handed helix are shown in (a). $\vb{\hat{N}}$ and $\vb{\hat{B}}$ lie in the plane which is normal to the tangent vector $\vb{\hat{T}}$ of the helical path. When defining a circle in the NB plane (green with red border) and extending it infinitely along the third direction, a Frenet-Serret waveguide is created. An orthographic side-view of the waveguide is depicted in (b). Red stripes show the orientation of the circular cross sections.}	
	\label{G_FrenetWaveguide}
\end{figure}

\subsection{Helicoidal Coordinate System}

Another local coordinate system to describe structures which are invariant along a helical path is the helicoidal frame. The helicoidal coordinates $(\xi_1,\xi_2,\xi_3)$ are related to the Cartesian coordinates $(x,y,z)$ via \cite{Russell2017}:
\begin{align} \label{TrafoCoordInv}
	\vb{r} = 
	\begin{pmatrix}
		x \\
		y \\
		z
	\end{pmatrix}
	=
	\begin{pmatrix}
		\xi_1 \cos(\alpha \xi_3) + \xi_2 \sin(\alpha \xi_3) \\
		-\xi_1 \sin(\alpha \xi_3) + \xi_2 \cos(\alpha \xi_3) \\
		\xi_3
	\end{pmatrix} &&
	\Leftrightarrow &&
	\begin{pmatrix}
		\xi_1 \\
		\xi_2 \\
		\xi_3
	\end{pmatrix}
	=
	\begin{pmatrix}
		x \cos(\alpha z) - y \sin(\alpha z) \\
		x \sin(\alpha z) + y \cos(\alpha z) \\
		z
	\end{pmatrix}.
\end{align}
For fixed values $\xi_1$ and $\xi_2$, the curve $\vb{r}(\xi_3)$ is a left-handed helix for $\alpha > 0$ (see Table \ref{HelixDef}). From this definition, the basis vectors of the helicoidal frame can be derived:
\begin{align}
	\begin{split}
		\vb{\xi_1} =
		\frac{\partial \vb{r}}{\partial \xi_1} =
		\begin{pmatrix}
			\cos(\alpha \xi_3) \\
			- \sin(\alpha \xi_3) \\
			0
		\end{pmatrix}, 
		\hspace{35pt}
		\vb{\xi_2} =
		\frac{\partial \vb{r}}{\partial \xi_2} =
		\begin{pmatrix}
			\sin(\alpha \xi_3) \\
			\cos(\alpha \xi_3) \\
			0
		\end{pmatrix},  \\
		\vb{\xi_3} =
		\frac{\partial \vb{r}}{\partial \xi_3} = 
		\begin{pmatrix}
			-\xi_1 \alpha \sin(\alpha \xi_3) + \xi_2 \alpha \cos(\alpha \xi_3) \\
			-\xi_1 \alpha \cos(\alpha \xi_3) - \xi_2 \alpha \sin(\alpha \xi_3) \\
			1
		\end{pmatrix}. 	 \hspace{35pt}
	\end{split}
\end{align}
Note, that $\vb{\xi_3}$ is not normalized and the system $(\vb{\xi_1},\vb{\xi_2},\vb{\xi_3})$ is not orthogonal. Since $\vb{\xi_1}$ and $\vb{\xi_2}$ always lie in the $xy$ plane, the helicoidal coordinate system is especially useful if the wavefronts of the fundamental mode are perpendicular to the $z$ axis. This is usually true for on-axis twisted structures.  \\ 

\begin{figure}[h!]
	\centering
	\includegraphics[scale=1]{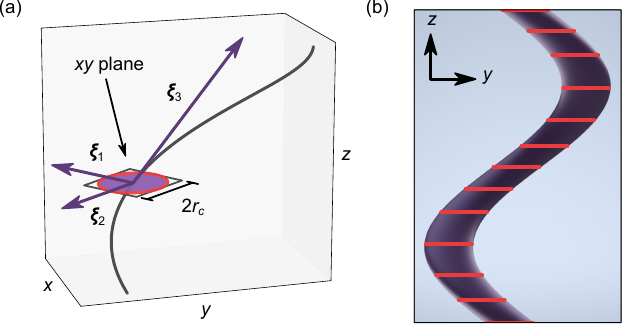}
	\caption[]{Definition of a helical waveguide in the helicoidal frame. The basis vectors of the helicoidal frame for a left-handed helix are shown in (a). Note, that only $\vb{\xi_1}$ and $\vb{\xi_2}$ are normalized, which simplifies calculations compared to a complete normalization of the system. When defining a circle in the $xy$ plane (purple with red border) and extending it infinitely along the $\vb{\xi_3}$ direction, a helicoidal waveguide is created. An orthographic side-view of the waveguide is depicted in (b). Red stripes show the orientation of the circular cross sections.}	
	\label{G_HelicoidalWaveguide}
\end{figure}

A helical waveguide can be constructed in this coordinate system by defining a permittivity profile $\epsilon$ in the $\xi_1\xi_2$ plane (which is identical to the $xy$ plane) and extending it infinitely along the $\xi_3$ coordinate:
\begin{equation} \label{EpsilonTwist}
	\begin{gathered}
		\epsilon (\xi_1,\xi_2,\xi_3)  = \epsilon (\xi_1,\xi_2, 0)  \\ \Longleftrightarrow \\
		\epsilon (x,y,z) = \epsilon (x \cos(\alpha z) - y \sin(\alpha z), x \sin(\alpha z) + y \cos(\alpha z), 0).
	\end{gathered}
\end{equation}
When this profile is defined to be a circle, we refer to the resulting structure as a "helicoidal waveguide". Note, that the helicoidal waveguide differs from the Frenet-Serret waveguide in the way the circles are oriented (cf. Figs.\ \ref{G_FrenetWaveguide}, \ref{G_HelicoidalWaveguide}).

\subsection{Overfelt Coordinate System}

A third coordinate system in which a helical structure is translationally invariant along one axis is a coordinate frame $(\rho, \phi, \zeta)$ used by Overfelt \cite{Overfelt2001}. Its relation to the Cartesian lab frame is similar to the definition of cylindrical coordinates but includes an additional dependence on the angular twist rate $\alpha$:
\begin{align} \label{TrafoOverfelt}
	\vb{r}= 
	\begin{pmatrix}
		x \\
		y \\
		z
	\end{pmatrix}
	=
	\begin{pmatrix}
		\rho \cos(\phi) \\
		\rho \sin(\phi) \\
		\zeta + \phi/\alpha 
	\end{pmatrix} &&
	\Longleftrightarrow &&	
	\begin{pmatrix}
		\rho \\
		\phi \\
		\zeta
	\end{pmatrix}
	=
	\begin{pmatrix}
		\sqrt{x^2+y^2} \\
		\tan^{-1}(y/x) \\
		z- \tan^{-1}(y/x)/\alpha 
	\end{pmatrix}.
\end{align}
When fixing a particular value of $\rho$ and $\zeta$, $\vb{r}(\phi)$ describes a right-handed helix for $\alpha > 0$. By contrast, the helicoidal coordinate system describes a left-handed helix for $\alpha>0$. The basis vectors of the Overfelt system $(\vb{e_\rho},\vb{e_\phi},\vb{e_\zeta})$ are nonorthogonal and can be calculated as:
\begin{align}
	\vb{e_\rho} =
	\frac{\partial \vb{r}}{\partial \rho} =
	\begin{pmatrix}
		\cos(\phi) \\
		\sin(\phi) \\
		0
	\end{pmatrix}, &&
	\vb{e_\phi} =
	\frac{\partial \vb{r}}{\partial \phi} = \frac{1}{\alpha}
	\begin{pmatrix}
		- \alpha \rho \sin(\phi) \\
		\alpha \rho \cos(\phi) \\
		1
	\end{pmatrix}, &&
	\vb{e_\zeta} =
	\frac{\partial \vb{r}}{\partial \zeta} =
	\begin{pmatrix}
		0 \\
		0 \\
		1
	\end{pmatrix}.
\end{align}
To construct a helical waveguide, a refractive index profile is defined in the $\rho \zeta$ plane (which is identical to the $\rho z$ plane of the cylindrical coordinate system) and extended infinitely along the $\phi$ coordinate. When this profile is a circle, we refer to the resulting structure as an "Overfelt waveguide".
\begin{figure}[h!]
	\centering
	\includegraphics[scale=1]{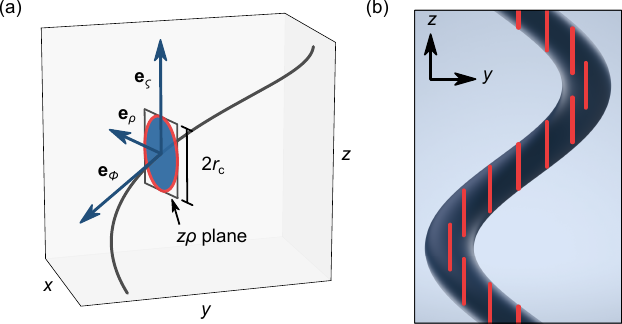}
	\caption[]{Definition of a helical waveguide in the Overfelt frame. The normalized basis vectors of the Overfelt frame for a left-handed helix are shown in (a). When defining a circle in the $z \rho$ plane (blue with red border) and extending it infinitely along the $\vb{e_\phi}$ direction, an Overfelt waveguide is created. An orthographic side-view of the waveguide is depicted in (b). Red stripes show the orientation of the circular cross sections.}	
	\label{G_OverfeltWaveguide}
\end{figure}

\section{Cross section of the three investigated waveguide types in the xy plane} \label{SI_CrossSecFormula}

Knowing the cross sections of the three waveguides in the $xy$ plane allows to perform all simulations in the helicoidal coordinate system because the shape of the cross sections merely rotates around the twist axis as the $z$ coordinate is changed. \\

\begin{figure}[h!]
	\centering
	\includegraphics[scale=1]{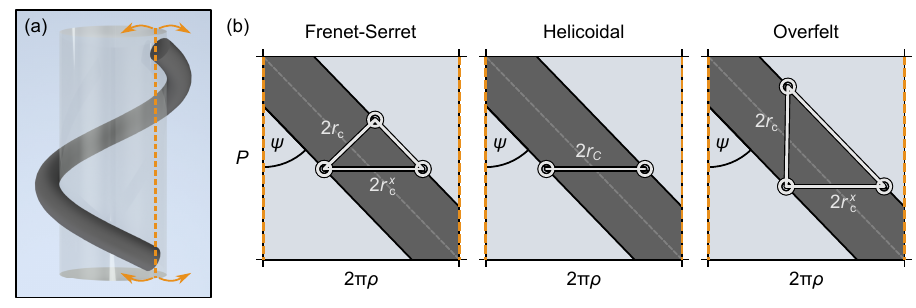}
	\caption[]{Approximation for the $xy$ cross section of the investigated helical waveguides as ellipses. When imagining the waveguides to be placed on the surface of a cylinder (a), the surface can be cut along the orange dashed line and unrolled onto a plane (b), provided that the waveguides are weakly coiled ($\bar{\kappa} r_c \ll 1$). The waveguides are tilted by an angle $\psi$, giving rise to elliptical cross sections in the $xy$ plane with semi-axes $r_c^x$ and $r_c^y$, as long as $2 r_c^x \ll 2 \pi \rho$. $r_c^y$ points into the plane and is equal to $r_c$ for all three waveguides. In its defining coordinate systems, each waveguide has a circular cross section with radius $r_c$.}	
	\label{fgr:CylinderHelixUnroll}
\end{figure}

An analytical description of the waveguides' cross sections is possible using an idea adapted from Ref. \cite{Russell2017}: When imagining the helical waveguide to be wound around a cylinder, it is possible to 'cut' the cylinder along the axial direction and unroll it onto a plane as shown in Fig.\ \ref{fgr:CylinderHelixUnroll}. We point out that this unrolling is only possible in the limit of weak coiling, i.e., if the radius of curvature of the helix is much larger than the radius of the core $(\bar{\kappa} r_c \ll 1)$. Otherwise, the side of the waveguide facing away from the cylinder is much longer than the side facing towards the inside preventing the unrolling. \\

Noting that the waveguides are tilted by an angle $\psi = tan^{-1}(\alpha \rho)$ from the $z$ axis after unrolling, it is possible to describe the cross sections of the waveguides in the $xy$ plane by an ellipse. One axis of the ellipse is perpendicular to the surface of the cylinder and its length $r_c^y$ is equal to the radius $r_c$ of the circular cross section defined in the corresponding coordinate system. The length of the other axis $r_c^x$ can be calculated by trigonometry from the sketches in Fig.\ \ref{fgr:CylinderHelixUnroll}(b), yielding the values shown here in Table \ref{ParamsEllipses}.
\begin{table}[h!]
	\centering
	\caption{\label{ParamsEllipses} Semi-major and semi-minor axes of the elliptical $xy$ cross sections of the three helical waveguides in the weak-coiling approximation, and provided that $2 r_c^x \ll 2 \pi \rho$. Lengths of the semi-axes of the ellipses are denoted as $r_c^x$ and $r_c^y$. $r_c$ is the radius of the core defined within each of the coordinate systems.}
	\begin{tabular}{cccc}
		\hline \hline
		\rule{0pt}{2ex}
		 \ \ Semi-axis \ \ &  \ \ Frenet-Serret  \ \ &  \ \ Helicoidal  \ \ &  \ \ Overfelt  \ \ \\[1mm]
		\hline
		\rule{0pt}{3ex}
		$r_c^x$   & $\sqrt{1+(\alpha \rho)^2} \, r_c $  & $r_c$ & $\alpha \rho \, r_c $ \\
		$r_c^y$   & $r_c$ & $r_c$ & $r_c$ \\
		\hline \hline
	\end{tabular}
\end{table}

To be precise, the so-obtained cross section would now have to be "attached" back onto the cylinder which deforms the ellipse according to the curvature of the cylinder (see right panels in Fig.\ 2 of the main text). However, this step would generate a more difficult shape and is not required as long as the length of the ellipse along the azimuthal direction (here $2 r_c^x$) is much smaller than the circumference ($2 r_c^x \ll 2 \pi \rho$).

\section{Transformation of fields from helicoidal frame to lab frame} \label{SI_TransformationGeneral}

All simulations were performed using the helicoidal coordinate system, as implemented in a commercial FEM solver (PropagatingMode module of JCMwave). Due to the invariance of the waveguide along the $\xi_3$ coordinate, eigenmodes of helical waveguides can be separated into a phase factor and a modal field, which does not depend on $\xi_3$ \cite{Weiss2013}:
\begin{equation} \label{Separation}
	\vb{\tilde{F}(\xi_1,\xi_2,\xi_3)} = \mathrm{e}^{\mathrm{i} \kappa \xi_3} \vb{F(\xi_1,\xi_2)},
\end{equation}
where $\kappa/k_0$ is the eigenvalue of the mode returned by the mode solver. The resulting field needs to be transformed back to the Cartesian lab frame as described in detail in this section. Furthermore, we show under which conditions it is possible to define an effective mode index in the lab frame.

We assume that the solution of the field $\vb{E}$ is known for the plane $z=\xi_3=0$ and the field at an arbitrary point $(\tilde{x},\tilde{y},\tilde{z})$ is to be calculated in the lab frame. The field components are transformed from the helicoidal frame to the lab frame by the Jacobian of the coordinate transformation $\underline{\vb{T}}$:
\begin{equation}
	\vb{E} = E_x \vb{\hat{x}} + E_y \vb{\hat{y}} + E_z \vb{\hat{z}} = E_{\xi_1} \vb{\xi_1} + E_{\xi_2} \vb{\xi_2} + E_{\xi_3} \vb{\xi_3}\, , \ \ \ \  \mathrm{with}	\ \ \ \ 
	\underbrace{	\begin{pmatrix}
			E_x \\
			E_y \\
			E_z
	\end{pmatrix}}_{\vb{E}^{\mathrm{Lab}}}
	= \underline{\vb{T}} 
	\underbrace{	\begin{pmatrix}
			E_{\xi_1} \\
			E_{\xi_2}  \\
			E_{\xi_3} 
	\end{pmatrix}}_{\vb{E}^{\mathrm{Helical}}}.
\end{equation}
$\vb{E}_{\mathrm{Helical}}$ and $\vb{E}_{\mathrm{Lab}}$ were introduced to denote the coordinate vectors of $\vb{E}$ in the two frames. For the helicoidal coordinate system, the Jacobian and its inverse $\underline{\vb{T}}^{-1}$ read:
\begin{equation} \label{Trafo}
		\underline{\vb{T}} = (\vb{\xi_1},\vb{\xi_2},\vb{\xi_3}) =
		\begin{pmatrix}
			\cos(\alpha \xi_3) & \sin(\alpha \xi_3) & \alpha [-\xi_1 \sin(\alpha \xi_3) + \xi_2 \cos(\alpha \xi_3)] \\
			-\sin(\alpha \xi_3) & \cos(\alpha \xi_3) & 		-\alpha [\xi_1 \cos(\alpha \xi_3) + \xi_2 \sin(\alpha \xi_3)] \\ 
			0 & 0 & 1
		\end{pmatrix} \\ 
		=
		\begin{pmatrix}
			\cos(\alpha z) & \sin(\alpha z) & \alpha y \\
			-\sin(\alpha z) & \cos(\alpha z) & 		- \alpha x \\ 
			0 & 0 & 1
		\end{pmatrix},
\end{equation}

\begin{equation} \label{TrafoInv}
		\underline{\vb{T}}^{-1} =
		\begin{pmatrix}
			\cos(\alpha \xi_3) & -\sin(\alpha \xi_3) & -\alpha \xi_2   \\
			\sin(\alpha \xi_3) & \cos(\alpha \xi_3) & 		\alpha \xi_1 \\ 
			0 & 0 & 1
		\end{pmatrix} \\ 
		=
		\begin{pmatrix}
			\cos(\alpha z) & -\sin(\alpha z) & -\alpha [x \sin(\alpha z) + y \cos(\alpha z)] \\
			\sin(\alpha z) & \cos(\alpha z) & 		\alpha [x \cos(\alpha z) -y \sin(\alpha z)]  \\ 
			0 & 0 & 1
		\end{pmatrix}.
\end{equation}
Furthermore, it is useful to define the transformation matrices at the initial plane $(\xi_3=z=0)$:
\begin{equation} \label{Trafo0Inv}
	\underline{\vb{T_0}} =
	\begin{pmatrix}
		1 & 0 & \alpha \xi_2  \\
		0 & 1 & 		-\alpha x\xi_1  \\ 
		0 & 0 & 1
	\end{pmatrix}, \ \ \ \ 
	\underline{\vb{T_0}}^{-1} =
	\begin{pmatrix}
		1 & 0 & -\alpha \xi_2  \\
		0 & 1 & 		\alpha \xi_1  \\ 
		0 & 0 & 1
	\end{pmatrix}.
\end{equation}
\newpage
In practice, the transformation between the coordinate systems is performed using the following steps:
\begin{enumerate}
	
	\item The point $(\tilde{x},\tilde{y},\tilde{z})$ in the lab frame corresponds to a certain point $(\tilde{\xi_1},\tilde{\xi_2},\tilde{\xi_3})$ in the helocoidal frame, which is calculated by Eq.\ (\ref{TrafoCoordInv}):
	\begin{equation}
		\begin{pmatrix}
			\tilde{\xi_1} \\
			\tilde{\xi_2} \\
			\tilde{\xi_3}
		\end{pmatrix}
		=
		\begin{pmatrix}
			\tilde{x} \cos(\alpha \tilde{z}) - \tilde{y} \sin(\alpha \tilde{z}) \\
			\tilde{x} \sin(\alpha \tilde{z}) + \tilde{y} \cos(\alpha \tilde{z}) \\
			\tilde{z}
		\end{pmatrix}.
	\end{equation}	
	\item Next, the field components in the helical frame $\vb{E}^{\mathrm{Helical}}$ at the point $(\tilde{\xi_1},\tilde{\xi_2},\xi_3=0)$ are determined. Since JCMwave returns the field components in the lab frame $\vb{E}^{\mathrm{Lab}}$ at $z=0$, they need to be converted to the helical frame keeping $z=\xi_3=0$ using Eq.\ (\ref{Trafo0Inv}):
	\begin{equation}
		\vb{E}^{\mathrm{Helical}}(\tilde{\xi_1},\tilde{\xi_2},\xi_3=0)	= \underline{\vb{T_0}}^{-1}  \vb{E}^{\mathrm{Lab}}(\tilde{x} \cos(\alpha \tilde{z}) - \tilde{y} \sin(\alpha \tilde{z}), \tilde{x} \sin(\alpha \tilde{z}) + \tilde{y} \cos(\alpha \tilde{z}),z=0).
	\end{equation}	
	Here, we used that $\xi_1=x$ and $\xi_2=y$ for $\xi_3=z=0$ according to Eq.\ (\ref{TrafoCoordInv}). Note, that the point at which the solution $\vb{E}^{\mathrm{Lab}}$ is evaluated depends on $z$ due to the rotation of the two coordinate frames in relation to each other. Other solvers might directly return the field components in the helicoidal frame, in which case this step can be omitted.
	\item The phase evolution factor from Eq.\ (\ref{Separation}) is introduced:
	\begin{equation}
		\vb{E}^{\mathrm{Helical}}(\tilde{\xi_1},\tilde{\xi_2},\tilde{\xi_3}) = \mathrm{e}^{\mathrm{i} \kappa \tilde{\xi_3}} \vb{E}^{\mathrm{Helical}}(\tilde{\xi_1},\tilde{\xi_2},\xi_3=0).	
	\end{equation}
	
	\item The field is transformed back to the lab frame using Eq.\ (\ref{Trafo}):
	\begin{equation} \label{FieldTrafoMain}
		\begin{split}
			&\vb{E}^{\mathrm{Lab}}(\tilde{x},\tilde{y},\tilde{z}) = \underline{\vb{T}} \vb{E}^{\mathrm{Helical}}(\tilde{\xi_1},\tilde{\xi_2},\tilde{\xi_3}) \\[0.2ex]
			&= 	\underbrace{\mathrm{e}^{\mathrm{i} \kappa \tilde{z}} \vphantom{\vb{E}^{\mathrm{Lab}}(\tilde{x} \cos(\alpha \tilde{z}) - \tilde{y} \sin(\alpha \tilde{z}), \tilde{x} \sin(\alpha \tilde{z}) + \tilde{y} \cos(\alpha \tilde{z}),z=0)}}_{\mathrm{phase}}
			\underbrace{	\begin{pmatrix}
					\cos(\alpha \tilde{z}) & \sin(\alpha \tilde{z}) & 0 \\
					-\sin(\alpha \tilde{z}) & \cos(\alpha \tilde{z}) & 	0 \\ 
					0 & 0 & 1
			\end{pmatrix}}_{\mathrm{rotation \ of \ field}}
			\vb{E}^{\mathrm{Lab}}(\underbrace{\tilde{x} \cos(\alpha \tilde{z}) - \tilde{y} \sin(\alpha \tilde{z}), \tilde{x} \sin(\alpha \tilde{z}) + \tilde{y} \cos(\alpha \tilde{z}),z=0}_{\mathrm{rotation \ of\ coordinates}}).
		\end{split}
	\end{equation}
	
\end{enumerate}
The solution contains three contributions: (1) the phase evolution calculated in the helicoidal frame, (2) a rotation of the polarization ellipse around the $z$ axis, and (3) the rotation of the coordinate frames with respect to each other. \\

An important observation is that the 'shape' of the 3D polarization ellipse remains the same as the field propagates along the waveguide and merely rotates around the $z$ axis. For fields where the polarization is the same at every point in the $xy$ plane, one can therefore state that:

\begin{enumerate}
	\item Transverse fields remain transverse:
	\begin{equation}
		E_{z} (x,y,0) = 0 \iff E_{z} (x,y,z) = 0.
	\end{equation}
	
	\item The degree of circularity as defined in Eq.\ (\ref{StokesS3Mag}) does not change:
	\begin{equation}
		\hat{S}_3 (x,y,0) = \hat{S}_3 (x,y,z).
	\end{equation}
	
\end{enumerate}

Eq.\ (\ref{FieldTrafoMain}) can be applied to calculate the phase evolution in the lab frame, which is important when studying mode coupling with structures that are not part of the twisted frame (e.g., a straight waveguide running parallel to a twisted waveguide). A necessary condition for mode coupling to occur is typically stated in the form that the effective index of the two modes needs to match. In the following subsections we show for a few relevant examples under which conditions it is possible to define an effective index in the lab frame - which is not always possible.

\subsection{Fields located on-axis}

\begin{figure}[h!] 
	\centering
	\includegraphics[scale=1]{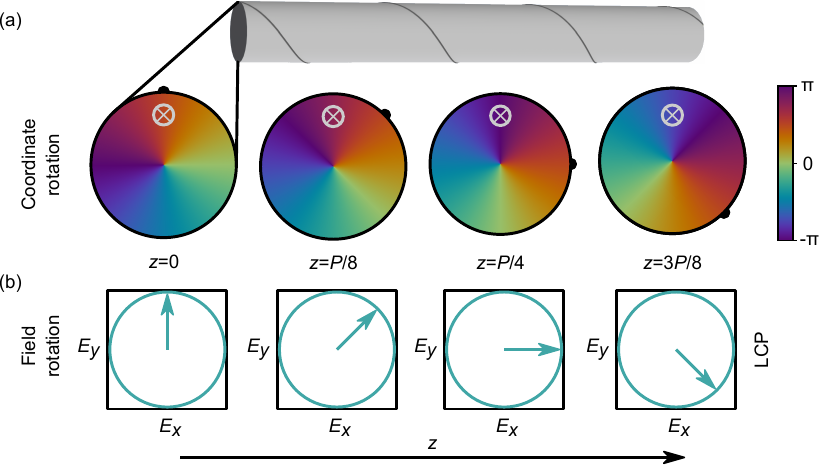} 
	\caption[]{Graphical explanation of the transformation phase for a circularly polarized transverse field with s=l=1. For on-axis fields, the phase evolution is evaluated at fixed point $(x,y)$ in the lab frame (gray crossed circle) as the $z$ coordinate is increased. (a) shows the consequence of the rotation of the helical coordinate frame with respect to the lab frame. To isolate this effect, the phase of the $x$ component of $\vb{E}^{\mathrm{Lab}}(x \cos(\alpha z) - y \sin(\alpha z), x \sin(\alpha z) + y \cos(\alpha z),0)$ is evaluated [right hand side of Eq.\ (\ref{FieldTrafoMain})]. As $z$ increases, the phase at $(x,y)$ grows resulting in the OAM contribution to the transformation phase: $l \alpha z$. The black bulge is a guide to the eye. (b) shows the isolated effect of the rotation of the field [middle term in Eq.\ (\ref{FieldTrafoMain})]. As $z$ increases the polarization ellipse is rotated with a period equal to the pitch length $P$ of the helicoidal coordinate system, resulting in the spin contribution to the transformation phase: $s \alpha z$. Note, that for the structure shown here, the sum of transformation phase and helical propagation phase does not depend on the twist rate due to its circular symmetry.}		
	\label{CoordFieldRotation}
\end{figure}

Although not part of this work, we start by analyzing the properties of the transformation for twisted waveguides where fields are located on-axis. Such waveguides have attracted increasing interest over the last decade, in particular twisted hollow-core and twisted photonic crystal fibers \cite{Russell2017, Chen2021, Edavalath2017, Roth2018, Roth2019, Wong2012, Xi2014}, as well as on-axis twisted step-index fibers \cite{Ren2021}. For on-axis fields it is possible to track the phase evolution of a fixed point $(x,y)$ in the lab frame as the $z$ coordinate is increased to check whether an effective index can be defined (see Fig.\ \ref{CoordFieldRotation}).

\subsubsection{Circularly polarized beams with OAM phase profile} \label{TransformationPhaseFirstMention}

The simplest case is a transverse field in the lab frame which is perfectly circularly polarized at every point in space at $z=0$ and has an OAM phase profile with the singularity located on the twist axis:
\begin{equation} \label{InputSOAMField}
	\vb{E}_{s,l}^{\mathrm{Lab}}(x,y,z=0) = \frac{1}{\sqrt{2}} \begin{pmatrix}
		1 \\
		s \mathrm{i}  \\
		0 
	\end{pmatrix}
	\mathrm{e}^{\mathrm{i} l \phi},
\end{equation}
where $s=\pm 1$ denotes the spin, $l \in \mathbb{Z}$ the OAM order and $\phi = \tan^{-1}(y/x)$ the azimuthal angle. The field in the lab frame at $z\neq 0$ can be obtained by applying Eq.\ (\ref{FieldTrafoMain}), yielding:
\begin{equation} 
	\vb{E}_{s,l}^{\mathrm{Lab}}(x,y,z) = \frac{1}{\sqrt{2}} \begin{pmatrix}
		1 \\
		s \mathrm{i}  \\
		0 
	\end{pmatrix}
	\mathrm{e}^{\mathrm{i} l \tan^{-1}\left(\frac{x \sin(\alpha z) + y \cos(\alpha z)}{x \cos(\alpha z) - y \sin(\alpha z)}\right)}  \mathrm{e}^{\mathrm{i} s \alpha z} \mathrm{e}^{\mathrm{i} \kappa z}.
\end{equation}
Using the addition theorem for tangens, $\tan (\gamma + \delta) = \frac{\tan(\gamma)+tan(\delta)}{1-\tan(\gamma) tan(\delta)}$, the term in the exponential can be simplified as:
\begin{equation}	
	\tan^{-1}\left(\frac{x \sin(\alpha z) + y \cos(\alpha z)}{x \cos(\alpha z) - y \sin(\alpha z)}\right)
	= \tan^{-1}\underbrace{\left(\frac{\tan(\alpha z) + y/x}{1 - \tan(\alpha z) y/x}\right)}_{\tan[\alpha z+ \tan^{-1}(y/x)]}
	= \alpha z + \tan^{-1}(y/x).
\end{equation}
The resulting equation for the evolution of the field in the lab frame is therefore:
\begin{equation} \label{OAMCircTrafo}
	\vb{E}_{s,l}^{\mathrm{Lab}}(x,y,z) = \frac{1}{\sqrt{2}} \begin{pmatrix}
		1 \\
		s \mathrm{i}  \\
		0 
	\end{pmatrix}
	\mathrm{e}^{\mathrm{i} l \phi} \mathrm{e}^{\mathrm{i} (s+l) \alpha z} \mathrm{e}^{\mathrm{i} \kappa z}
	=  \vb{E}_{s,l}^{\mathrm{Lab}}(x,y,z=0)  \mathrm{e}^{\mathrm{i} (s+l) \alpha z} \mathrm{e}^{\mathrm{i} \kappa z}.
\end{equation}

We denote $\kappa z$ as the \textbf{helical propagation phase} and $(s+l) \alpha z$ as the \textbf{transformation phase}. For this example, both phases increase linearly in $z$ which allows to define an effective index in the lab frame $n_{\mathrm{eff}}^{\mathrm{Lab}}$ as:
\begin{equation}	\label{EffIndexTrafo}
	n_{\mathrm{eff}}^{\mathrm{Lab}} = n_{\mathrm{eff}}^{\mathrm{Helical}} + (s+l) \frac{\alpha \lambda}{2 \pi},
\end{equation}
where $n_{\mathrm{eff}}^{\mathrm{Helical}}(\alpha) = \kappa/k_0$ is the effective index in the helical frame. Note, that this result does not contain approximations and holds for arbitrary twist rates as long as the fields have the structure defined in Eq.\ (\ref{InputSOAMField}).

The origin of the transformation phase is illustrated in Fig.\ \ref{CoordFieldRotation} for a circularly symmetric structure. According to Eq.\ (\ref{FieldTrafoMain}), a field that is invariant under changes of $\xi_3$ in the helicoidal frame rotates in the lab frame as $z$ is increased, following the helical path of the structure. If the field carries OAM, the rotation of this phase profile results in the OAM contribution to the transformation phase $l \alpha z$. Eq.\ (\ref{FieldTrafoMain}) shows that apart from the phase profile, the polarization ellipse is also rotated when $z$ increases. This rotation results in the spin contribution to the transformation phase $s \alpha z$. \\ 

It is important to note, that the helical propagation phase typically also carries a dependence on $\alpha$. For the structure shown in Fig.\ \ref{CoordFieldRotation}, it is straightforward to see that the transformation phase is exactly canceled by the twist rate dependence of the helical propagation phase: An on-axis structure with circular symmetry cannot be twisted (i.e., it does not change when twisted). For systems which are not circularly symmetric (e.g., hexagonal cores), the sum of the two phases remains dependent on the twist rate giving rise to circular birefringence.

\subsubsection{Elliptically polarized beams}

Next, we study an example of eigenmodes which are transverse and elliptically polarized. For simplicity, the phase profile is assumed to be flat and the polarization is set to be constant across the beam cross section for $z=0$. 
\begin{equation} \label{InputEllField}
	\vb{E}_{\tilde{s},l}^{\mathrm{Lab}}(x,y,z=0) = \frac{1}{\sqrt{1+(1-\epsilon)^2}} \begin{pmatrix}
		1 \\
		\tilde{s} (1-\epsilon) \mathrm{i}  \\
		0 
	\end{pmatrix},
\end{equation}
where $\tilde{s}=\pm 1$ determines the direction of rotation of the polarization vector. Different values of the ellipticity of the polarization ellipse are investigated, which is defined as \cite{Okamoto2006}:
\begin{equation}
	\epsilon = \frac{a_x-a_y}{a_x},
\end{equation}
where $a_x$ denotes the length of the major axis of the polarization ellipse and $a_y$ the length of the minor axis. $\epsilon = 0$ corresponds to circular polarization and $\epsilon = 1$ to linear polarization. Applying Eq.\ (\ref{FieldTrafoMain}), yields the field at $z \neq 0$:
\begin{equation}
	\vb{E}_{\tilde{s},l}^{\mathrm{Lab}}(x,y,z) = \frac{1}{\sqrt{1+(1-\epsilon)^2}} \underbrace{\begin{pmatrix}
			\cos(\alpha z)+\tilde{s} \sin(\alpha z) (1-\epsilon) \mathrm{i} \\
			- \sin(\alpha z) + \tilde{s} \cos(\alpha z) (1-\epsilon) \mathrm{i} \\
			0 
	\end{pmatrix}}_{\vb{E}^{\mathrm{Transformation}}} \mathrm{e}^{\mathrm{i} \kappa z}.
\end{equation}
While the helical propagation phase $\kappa z$ occurs in the same form as for circularly polarized fields, the transformation phase cannot be stated in a simplified form. Therefore, the transformation phase is determined numerically as the argument of $\vb{E}^{\mathrm{Transformation}}$. The increase of the transformation phase with $z$ is shown for several values of $\epsilon$ below.
\begin{figure}[h!]
	\centering
	\includegraphics[scale=1]{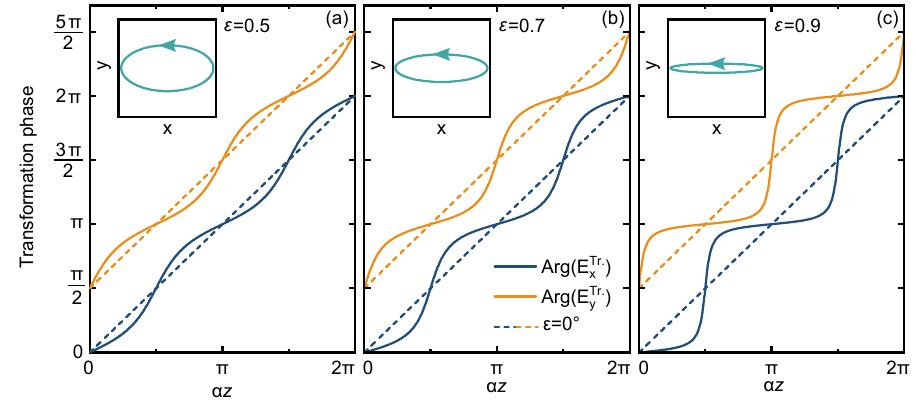}
	\caption[]{Transformation phase for elliptically polarized transverse fields with $\tilde{s}=1$. The phases of the $x$ and $y$ component of $\vb{E}^{\mathrm{Transformation}}$ are shown as solid blue and orange lines, respectively. As the ellipticity of the polarization ellipse $\epsilon$ is increased, the differences to the phases of circular polarization (shown as dashed lines) grow larger. For ellipticities of $\epsilon=0.5$ (a) and $\epsilon=0.7$ (b), the phase difference to LCP ($\epsilon=0$) remains below $0.2 \pi$. For a very high ellipticity of $\epsilon=0.9$ (c) a strong deviation from the linear increase in $z$ can be observed. Insets show the shape of the polarization ellipse. }	
	\label{EllPol}
\end{figure}

In untwisted waveguides, the propagation phase $\varphi$ always increases linearly in $z$, even if the waveguide is filled with an anisotropic material due to the invariance of the structure in the $z$ direction. This linearity is the basis for the definition of the effective refractive index $n_{\mathrm{eff}}$ via $\varphi = n_{\mathrm{eff}} k_0 z$. Twisted waveguides are however not invariant in the $z$ direction of the lab frame. A peculiar consequence is that the propagation phase of non-circularly polarized eigenmodes of twisted waveguides can differ substantially from this linear increase such as in Fig.\ \ref{EllPol}(c). Therefore, it is generally not advisable to define an effective index in the lab frame via Eq.\ (\ref{EffIndexTrafo}). For example, a situation could arise where an eigenmode of a twisted waveguide would be assigned the same effective index as an eigenmode of an untwisted waveguide although their phases do not match locally at each $z$ position. \\ 

More generally, the deviation from the linear increase in phase is a consequence of the rotation of the polarization ellipse as the field propagates along the twisted structure. For example, after propagating a quarter period of the helix pitch the polarization ellipse shown in Fig.\ \ref{EllPol}(c) would be oriented vertically instead of horizontally. This in a fundamental difference to fields propagating in untwisted waveguides, where the orientation of the polarization ellipse does not change. In this example, the overlap between the mode of a twisted and an untwisted waveguide would constantly change along the $z$ direction, corroborating the fact that modes of structures with different twist rates cannot simply be compared by looking at the effective index. \\

Nonetheless, Eq.\ (\ref{EffIndexTrafo}) still correctly predicts the average phase increase over one twist period (dashed lines in Fig.\ \ref{EllPol}) and can therefore serve as a guidance if the ellipticity of the field remains small enough such that the $z$ dependence of the overlap integral can be neglected. In that case the matching of effective indices defined by Eq.\ (\ref{EffIndexTrafo}) would give a necessary condition for quasi-phase-matching of two modes - similar to quasi-phase-matching in nonlinear optics.

\subsubsection{Beams with arbitrary phase profiles}

A similar deviation from the linear phase increase occurs if the phase profile of the beam does not depend exactly linearly on $\phi$, e.g., if the OAM phase profile is distorted due to coupling to other modes. For simplicity we study a fictitious example, where the azimuthal phase increases only in one half of the beam and is constant in the other half. The field is assumed to be  transverse and circularly polarized:
\begin{empheq}[]{align}
	\vb{E}_{s}^{\mathrm{Lab}}(x,y,z=0) = \frac{1}{\sqrt{2}} \begin{pmatrix}
		1 \\
		s \mathrm{i}  \\
		0 
	\end{pmatrix}
	\mathrm{e}^{\mathrm{i} f[\phi(x,y)]}, &&
	f(\phi) = 	\begin{cases}
		2 \phi & 0 \le \phi < \pi \\
		0 & \pi \le \phi < 2 \pi 
	\end{cases},
\end{empheq}
where $s=\pm 1$ denotes the spin $\phi \in [0,2 \pi[$ is the azimuthal angle and $f(\phi)$ defines the phase profile. Again, we calculate the field for $z \neq 0$ using Eq.\ (\ref{FieldTrafoMain}), yielding:
\begin{equation}
	\vb{E}_{s}^{\mathrm{Lab}}(x,y,z)		
	= \frac{1}{\sqrt{2}} \underbrace{\begin{pmatrix}
			1 \\
			s \mathrm{i}  \\
			0 
		\end{pmatrix}
		\mathrm{e}^{\mathrm{i} f\{\phi[x \cos(\alpha z) - y \sin(\alpha z), x \sin(\alpha z) + y \cos(\alpha z)]\}} \mathrm{e}^{\mathrm{i} s \alpha z}}_{\vb{E}^{\mathrm{Transformation}}} \mathrm{e}^{\mathrm{i} \kappa z}.
\end{equation}
While the helical propagation phase $\kappa z$ occurs in the same form as in the previous examples, the transformation phase needs to be extracted as the argument of $\vb{E}^{\mathrm{Transformation}}$ shown below.
\begin{figure}[h!]
	\centering
	\includegraphics[scale=1]{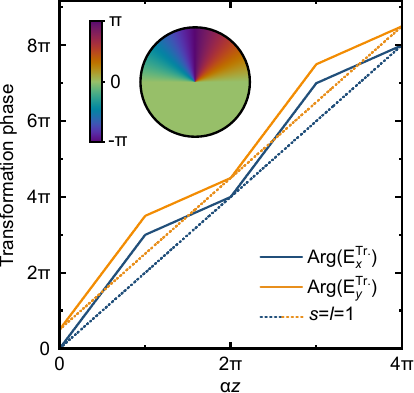}
	\caption[]{Transformation phase for a beam whose phase profile differs from that of an OAM beam. As an example, a fictitious mode with $s=1$ and the phase profile shown in the inset is assumed. The phases of the $x$ and $y$ component of $\vb{E}^{\mathrm{Transformation}}$ are shown as solid blue and orange lines, respectively. A strong deviation to the linear phase increase of a mode with perfect OAM phase profile with $s=l=1$ (dashed lines) can be observed.}	
	\label{NonOAMphase}
\end{figure}

Fig.\ \ref{NonOAMphase} confirms that care should be taken when using Eq.\ (\ref{EffIndexTrafo}) to determine an effective index in the lab frame for eigenmodes of twisted structures. If the spatial phase of modes differs substantially from $\mathrm{e}^{\mathrm{i} l \phi}$, the propagation phase of the mode does not increase linearly in the $z$ direction, which defeats the purpose of defining an effective index.

\subsection{Fields located off-axis} \label{OffAxisCore}

\begin{figure}[h!]
	\centering
	\includegraphics[scale=1]{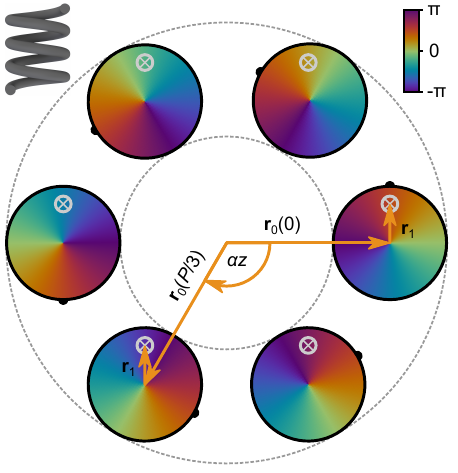}
	\caption[]{Graphical explanation of the OAM contribution to the transformation phase for fields located in an off-axis core. A circularly polarized transverse field with s=l=1 is shown in the lab frame as $z$ is increased in steps of 1/6 of the helix pitch. For off-axis fields, the phase evolution is evaluated at a fixed point $\vb{r}_1$ relative to the center of the moving core $\vb{r}_0(z)$. Note, that only the effect of the rotation of the helical coordinate frame with respect to the lab frame is shown. To isolate this effect, the phase of the $x$ component of $\vb{E}^{\mathrm{Lab}}(\tilde{x}
		\cos(\alpha z) - \tilde{y} \sin(\alpha z), \tilde{x} \sin(\alpha z) + \tilde{y} \cos(\alpha z),0)$ is evaluated [right hand side of Eq.\ (\ref{FieldTrafoMain})]. As $z$ increases, the phase at the tracked point (gray crossed circle) increases resulting in the OAM contribution to the transformation phase: $l \alpha z$. The black bulge is a guide to the eye.}	
	\label{GOffAxisCore}
\end{figure}

This subsection deals with fields located in off-axis waveguides, which are studied in this work. In this case it is not possible to track the phase evolution of a fixed point $(x,y)$ in the lab frame because as the $z$ coordinate is changed, the core that is carrying the field moves away from this position (see Fig.\ \ref{GOffAxisCore}) Therefore we track the phase evolution of a point that is located at a constant position relative to the center of the core as viewed in the lab frame.

Let's assume that the off-axis core is centered at a distance $\rho$ from the twist axis. In the lab frame this core moves along a helical trajectory $\vb{r}_0(z)$ according to Eq.\ (\ref{EpsilonTwist}). We now track a point that is placed at a constant (i.e., independent of $z$) offset $\vb{r}_1$ from this trajectory in the $xy$ plane with local coordinates $\bar{x}$ and $\bar{y}$. Without loss of generality, the position of the core at $z=0$ is chosen to be located on the positive part of the $x$ axis:
\begin{align}
	\vb{r}_0(z) = 	\begin{pmatrix}
		\rho \cos(\alpha z)  \\
		-\rho \sin(\alpha z)  \\
		z
	\end{pmatrix}, &&		
	\vb{r}_1 = \begin{pmatrix}
		\bar{x} \\
		\bar{y} \\
		0
	\end{pmatrix}, &&
	\vb{r}(\bar{x},\bar{y},z) = \vb{r}_0(z) + \vb{r}_1(\bar{x},\bar{y}) = \begin{pmatrix}
		\tilde{x}  \\
		\tilde{y}  \\
		\tilde{z}
	\end{pmatrix}.
\end{align}

\subsubsection{Circularly polarized beams with OAM phase profile}

As before, we first look at the simple case of a perfectly circularly polarized transverse field with a perfect OAM phase profile - now with the singularity located at the center of the off-axis core:
\begin{equation} \label{InputSOAMField_OffAxis}
	\vb{E}_{s,l}^{\mathrm{Lab}}(x,y,z=0) = \frac{1}{\sqrt{2}} \begin{pmatrix}
		1 \\
		s \mathrm{i}  \\
		0 
	\end{pmatrix}
	\mathrm{e}^{\mathrm{i} l \bar{\phi}},
\end{equation}
where $s=\pm 1$ denotes the spin, $l \in \mathbb{Z}$ the OAM order and $\bar{\phi} = \tan^{-1}[y/(x-\rho)]$ the azimuthal angle relative to $\vb{r}_0(z=0)$.

To calculate the field in the lab frame at $z\neq 0$, the field at $z=0$ needs to be evaluated according to \mbox{Eq.\ (\ref{FieldTrafoMain}):}
\begin{equation}
	\begin{split}
		& \vb{E}_{s,l}^{\mathrm{Lab}}(\tilde{x} \cos(\alpha \tilde{z}) - \tilde{y} \sin(\alpha \tilde{z}), \tilde{x} \sin(\alpha \tilde{z}) + \tilde{y} \cos(\alpha \tilde{z}),z=0) 	\\
		= \; &\vb{E}_{s,l}^{\mathrm{Lab}}(\rho + \bar{x} \cos(\alpha \tilde{z}) - \bar{y} \sin(\alpha \tilde{z}),\bar{x} \sin(\alpha \tilde{z}) + \bar{y} \cos(\alpha \tilde{z}) ,z=0) = \frac{1}{\sqrt{2}} \begin{pmatrix}
			1 \\
			s \mathrm{i}  \\
			0 
		\end{pmatrix}
		\mathrm{e}^{\mathrm{i} l \tan^{-1}\left(\frac{\bar{x} \sin(\alpha \tilde{z}) + \bar{y} \cos(\alpha \tilde{z})}{\bar{x} \cos(\alpha \tilde{z}) - \bar{y} \sin(\alpha \tilde{z})}\right)}.
	\end{split}
\end{equation}

Following the same algebra as in the previous section one arrives at:
\begin{equation} \label{OffAxisTrafo}
	\vb{E}_{s,l}^{\mathrm{Lab}}[\vb{r}(\bar{x},\bar{y},z)] = \frac{1}{\sqrt{2}} \begin{pmatrix}
		1 \\
		s \mathrm{i}  \\
		0 
	\end{pmatrix}
	\mathrm{e}^{\mathrm{i} l \tan^{-1}(\bar{y}/\bar{x})} \mathrm{e}^{\mathrm{i} (s+l) \alpha z} \mathrm{e}^{\mathrm{i} \kappa z}
	=  \vb{E}_{s,l}^{\mathrm{Lab}}[\vb{r}(\bar{x},\bar{y},z=0)]  \mathrm{e}^{\mathrm{i} (s+l) \alpha z} \mathrm{e}^{\mathrm{i} \kappa z}.
\end{equation}
This result has the same form as Eq.\ (\ref{OAMCircTrafo}) for on-axis fields. Therefore, we showed that it is also possible to define an effective index in the lab frame for off-axis fields - as long as the fields have the structure defined in Eq.\ (\ref{InputSOAMField_OffAxis}). The previously obtained formula for transforming the effective index Eq.\ (\ref{EffIndexTrafo}) also holds for these off-axis fields:
\begin{equation} \label{OffAxisConversion}
	n_{\mathrm{eff}}^{\mathrm{Lab}} = n_{\mathrm{eff}}^{\mathrm{Helical}} + (s+l) \frac{\alpha \lambda}{2 \pi}.
\end{equation}

\subsubsection{Non-transverse beams}

For a field in an off-axis core, its phase fronts are generally perpendicular to the helical curve on which the core moves. Therefore, the field develops a $z$ component and Eq.\ (\ref{OffAxisTrafo}) - derived for fields transverse to the $z$ axis - does not apply anymore. In this case, the field is usually still circularly polarized (i.e., $\abs{\hat{S}_3}=1$) but the 3D polarization ellipse does not lie in the $xy$ plane. \\

As an example, we study a beam with uniform circular polarization where the polarization ellipse is tilted out of the $xy$ plane by a tilt angle $\beta$:
\begin{equation} 
	\vb{E}_{s}^{\mathrm{Lab}}(x,y,z=0) = \frac{1}{\sqrt{2}} 
	\begin{pmatrix}
		1 & 0 & 0 \\
		0 & \cos(\beta) & 	-\sin(\beta) \\ 
		0 & \sin(\beta) & 	\cos(\beta)
	\end{pmatrix}		
	\begin{pmatrix}
		1 \\
		s \mathrm{i}  \\
		0 
	\end{pmatrix}	
	=  \frac{1}{\sqrt{2}} 				
	\begin{pmatrix}
		1 \\
		\mathrm{i} \cos(\beta)  \\
		\mathrm{i} \sin(\beta) 
	\end{pmatrix}.
\end{equation}
Again, Eq.\ (\ref{FieldTrafoMain}) is applied to calculate the field in the lab frame at $z\neq 0$. Since the field does not have a spatial dependence, the result is independent of the chosen trajectory $\vb{r}(x,y,z)$ and applies to both off-axis and on-axis fields:
\begin{equation} 
	\vb{E}_{s}^{\mathrm{Lab}}(x,y,z) 
	=  \frac{1}{\sqrt{2}} 	
	\underbrace{\begin{pmatrix}
			\cos(\alpha z) + \mathrm{i} \cos(\beta) \sin(\alpha z)  \\
			-\sin(\alpha z) + \mathrm{i} \cos(\beta) \cos(\alpha z)  \\
			\mathrm{i} \sin(\beta) 
	\end{pmatrix}}_{\vb{E}^{\mathrm{Transformation}}}
	\mathrm{e}^{\mathrm{i} \kappa z}.
\end{equation}
The transformation phase is calculated as the argument of $\vb{E}^{\mathrm{Transformation}}$. Note, that the phase of the $z$ component of $\vb{E}^{\mathrm{Transformation}}$ does not change because the transformation includes a rotation of the polarization ellipse around the $z$ axis. 

\begin{figure}[h!]
	\centering
	\includegraphics[scale=1]{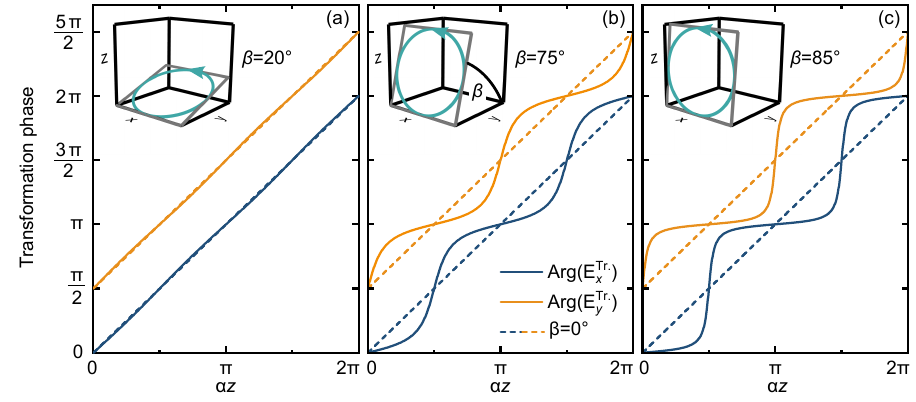}
	\caption[]{Transformation phase for circularly polarized fields with $s=1$ which are not transverse to the $z$ axis. The polarization ellipse is tilted out of the $xy$ plane by an angle $\beta$ as indicated in the figure. The phases of the $x$ and $y$ component of $\vb{E}^{\mathrm{Transformation}}$ are shown as solid blue and orange lines, respectively. As the tilt angle is increased, the differences to the phases of an untilted polarization ellipse (shown as dashed lines) grow larger. For tilt angles of $\beta=20 \degree$ (a) and $\beta=75 \degree$ (b), the phase difference to the untilted ellipse remains below $0.2 \pi$. Only for very high tilt angles of $\beta = 85 \degree$ (c) a strong deviation from the linear increase in $z$ can be observed. }	
	\label{TiltedEll}
\end{figure}

The transformation phase deviates from the linear increase in $z$ if the field is not transverse with respect to the $z$ axis. However, the phase difference to a transverse field with the same spin remains small even for high tilt angles of $\beta=75\degree$ [Fig.\ \ref{TiltedEll}(b)]. Therefore, Eq.\ (\ref{EffIndexTrafo}) can still be used as a guide to the phase evolution of the mode in the lab frame. 

\newpage
\subsection{Summary} \label{SummaryTrafo}

Eigenmodes of twisted waveguides have unique properties when it comes to their propagation. Since their field profile and their polarization ellipses rotate - following the twist of the structure - they can in general not be compared to a field propagating in an untwisted waveguide or a waveguide with a different twist rate. \\

However, an exception exists if the field fulfills the following conditions: 

\begin{itemize}
	\item the field is transverse to the $z$ axis 
	\item the field is circularly polarized with $s=\pm1$
	\item the spatial phase profile is flat or has an OAM profile with a $\mathrm{e}^{\mathrm{i} l \phi}$ phase dependence
\end{itemize}

Under these conditions, the propagation phase of the field increases linearly in the lab frame and an effective index in the lab frame can be defined using Eq.\ (\ref{EffIndexTrafo}):
\begin{equation}
	n_{\mathrm{eff}}^{\mathrm{Lab}} = n_{\mathrm{eff}}^{\mathrm{Helical}} + (s+l) \frac{\alpha \lambda}{2 \pi}.
\end{equation}
This equation holds both for on-axis fields and fields located in an off-axis twisted core. An effective index in the lab frame can still be defined if the conditions are only slightly violated, especially if the polarization is still circular but the polarization ellipse is tilted out of the $xy$ plane which is usually the case for off-axis twisted waveguides.

\section{Analytical description of Frenet-Serret Waveguides} \label{SI_Alexeyev}

Alexeyev and Yavorsky developed a thorough theoretical description of a Frenet-Serret waveguide \cite{Alexeyev2008}, which is used as a basis for this chapter. Their analysis is based on three assumptions: 

\begin{enumerate}
	\item The radius of curvature of the helix is small compared to the radius of the core: $r_c \bar{\kappa} \ll 1$ (weak coiling approximation)
	\item The fields are transverse in the Frenet-Serret frame and can be described by the scalar wave approximation. This approximation is valid for fibers with low index contrast: $\Delta = \frac{n_{\mathrm{Core}}^2-n_{\mathrm{Cladding}}^2}{2 n_{\mathrm{Core}}^2} \ll 1$.
	\item The torsion $\tau$ of the helix can be treated as a small parameter.
\end{enumerate}

Formulating the scalar wave equation in the Frenet-Serret frame introduces a new set of operators $\hat{V}_i$ compared to the operator $\hat{H}_0$ of the untwisted system:
\begin{equation}
	(\hat{H}_0 + \hat{V}_0 + \hat{V}_1 + ...) \ket{\Psi}_{\mathrm{Frenet}}= \beta_{\mathrm{Frenet}}^2 \ket{\Psi}_{\mathrm{Frenet}}.
\end{equation}
Here, $\ket{\Psi}_{\mathrm{Frenet}}$ denotes the two transverse components of the field and $\beta_{\mathrm{Frenet}}$ the propagation constant, both given in the Frenet-Serret frame. Expression for the operators can be found in \cite{Alexeyev2008}. Under the assumptions given above, the additional operators $\hat{V}_i$ can be treated as small perturbations of the untwisted system. Using first order perturbation theory, the twist-induced corrections to the propagation constant can be calculated. The dominant term is caused by the operator $\hat{V}_0$ and reads:
\begin{equation} \label{PhaseSplitting}
	\hspace{5pt}	\Delta \beta^{\mathrm{Frenet}} = (s+l) \tau \hspace{-22pt} \phantom{\text{\LARGE A}_\text{\large A}}.
\end{equation}
This simple result implies that any waveguide that forces photons to move along a non-planar path (which therefore has torsion) results in a twist-induced splitting of modes proportional to their total angular momentum $s+l$. For helical trajectories, the effect is particularly simple due to the constant torsion of a helix. \\

The result can be converted to the helicoidal coordinate frame by noting that the helix is parametrized by the arc length in the Frenet-Serret frame while it is parametrized by the $z$ distance in the helicoidal frame:
\begin{equation}
	n_{\mathrm{eff,Helicoidal}} = \left( n_0 +  \frac{(s+l) \tau}{k_0} \right) \sqrt{1+\alpha^2 \rho^2},
\end{equation}
where $n_0$ is the effective index of the mode in the untwisted waveguide. Note, that the sign of $\tau$ is negative for a left-handed helix. To get a better understanding of the magnitude of the twist-induced splitting, the result can further be converted to the lab frame using Eq.\ (\ref{OffAxisConversion}):
\begin{equation} \label{EffIndexLabFrame}
	n_{\mathrm{eff,Lab}} = n_0 \sqrt{1+\alpha^2 \rho^2} + (s+l) \frac{\alpha}{k_0} \left( 1- \frac{1}{\sqrt{1+\alpha^2 \rho^2}}  \right).
\end{equation}
This result is valid when the modal field is circularly polarized and has a flat or OAM phase profile (see Sec.\ \ref{SummaryTrafo}). The fact that the signs of $\tau$ and $\alpha$ are opposite was used.

For small twist rates $\alpha \rho \ll 1$ the result can be simplified showing that the splitting of modes with different total angular momentum increases strongly with twist rate and helix radius:
\begin{equation}
	n_{\mathrm{eff,Lab}} = n_0 \sqrt{1+\alpha^2 \rho^2} + (s+l) \frac{\rho^2 \alpha^3 \lambda}{4 \pi} \text{ \ \ \ \ \ for $\alpha \rho \ll 1$}.
\end{equation}
For high twist rates the splitting increases linearly in $\alpha$: 
\begin{equation}
	n_{\mathrm{eff,Lab}} = n_0 \sqrt{1+\alpha^2 \rho^2} + (s+l) \frac{\alpha \lambda}{2 \pi} \text{ \ \ \ \ \ for $\alpha \rho \gg 1$}.
\end{equation}
This important result shows that twisting a waveguide lifts the degeneracy between modes with the same magnitude of total angular momentum but different sign. It is remarkable that the twist-induced splitting is purely based on a geometric effect and is therefore not wavelength dependent \footnote{The phase difference $\delta n_{\mathrm{eff,Lab}} k_0 z$ does not depend on the wavelength.} Helical architectures therefore provide a novel way to create broadband spin- and OAM-preserving waveguides. More generally, any waveguide with nonzero torsion at each point of its trajectory will be able to preserve the angular momentum state of the light.

\subsection{Berry phase in helical waveguides} \label{BerrySpinOrbit}

On a more fundamental level, the additional phase induced by the twist [Eq.\ (\ref{PhaseSplitting})] can be seen as a consequence of the spin-orbit and orbit-orbit interaction of light in the "semi-geometrical optics" approximation (cf. Sec.\ \ref{SI_SHE}) \cite{Alexeyev2008}. As light propagates under the influence of these interactions, it acquires a spin- and OAM-dependent Berry phase $\Phi_B$ \cite{Chiao1986, Bliokh2015, Bliokh2006a}, the photonic equivalent of the Berry phase in quantum mechanics:
\begin{equation} \label{BerryPhaseEq}
	\Phi_B = (s+l) \int_{C} \vb{A} \cdot d\vb{k} = (s+l) \int_{S} \vb{F} \cdot d\vb{S} = (s+l) \Omega,
\end{equation}
where $C$ is the contour traced out by the light in momentum space, $\vb{F} = \grad_{\vb{k}} \cross \vb{A} = \frac{\vb{k}}{k^3} $ is the Berry curvature for $s=+1$ and $l=0$, and $\vb{A}$ is the Berry connection for $s=+1$ and $l=0$ (defined in Ref. \cite{Bliokh2015}). The second and third equalities hold if $C$ is a loop such that $S$ is its enclosed surface on the $k$-sphere ($\partial{S} = C$). In this case, the Berry phase is solely determined by the solid angle $\Omega$ of the loop that the light traces out on the $k$-sphere. \\

For off-axis twisted waveguides, $\vb{k}$ points along the tangent $\vb{\hat{T}}$ of the helix which is tilted from the $z$ axis by a fixed angle $\theta$ with $\cos(\theta) = 1/\sqrt{1+\alpha^2 \rho^2}$ (cf. Sec.\ \ref{HelixParamSec}). Therefore, the trajectory of the light traces out a solid angle $\Omega = 2 \pi [1-\cos(\theta)]$ after propagating one turn in the helix. Plugging these terms into Eq.\ (\ref{BerryPhaseEq}) yields the Berry phase shift after propagating a distance $z$ along the helix axis:
\begin{equation}
	\Phi_B =(s+l) \, 2 \pi \, \left(1-\frac{1}{\sqrt{1+\alpha^2 \rho^2}}\right) \, \frac{z}{P} \hspace{3mm} \mathrm{(Berry \ phase \ in \ a \ helical \ waveguide)}.
\end{equation}
The resulting correction to the effective index of the modes $\delta n$ can be calculated from $\delta n = \Phi_B/(z k_0)$ and coincides with the lab-frame result obtained earlier in Eq.\ (\ref{EffIndexLabFrame}).

\section{Fraction of power present inside the core of single-mode helical waveguides} \label{SI_FracPowerCore}

As explained in the main text in \rom{2} E 4, 
the geometry of the cross section of the helicoidal and Overfelt waveguide in the NB plane changes with twist rate. This strongly impacts the modes in the single-mode waveguides because more and more power is located in air outside of the core as the cross section becomes narrower. Since air has a lower refractive index than the core, this results in a reduction of the effective index.

\begin{figure}[h!]
	\centering
	\includegraphics[scale=1]{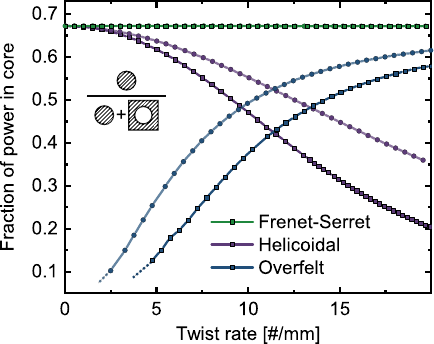}
	\caption[]{Twist rate dependence of the fraction of power present inside the core calculated for the single-mode helical waveguides geometries indicated in the legend. The power inside the core was calculated by integrating the T component of the Poynting vector over the area of the core in the NB plane. It was normalized by the integral over the complete simulation region as shown in the inset. A darker color shade denotes the LCP mode, a lighter shade the RCP mode. Note, that the modes of the Overfelt waveguide could not be calculated for twist rates below 5/mm due to very high loss and low confinement in the core (blue dots).}	
	\label{fgr:FracPowerCore_SI}
\end{figure}

\section{Superchiral fields in helical waveguides} \label{SI_Superchiral}

\begin{figure}[h!]
	\centering
	\includegraphics[scale=1]{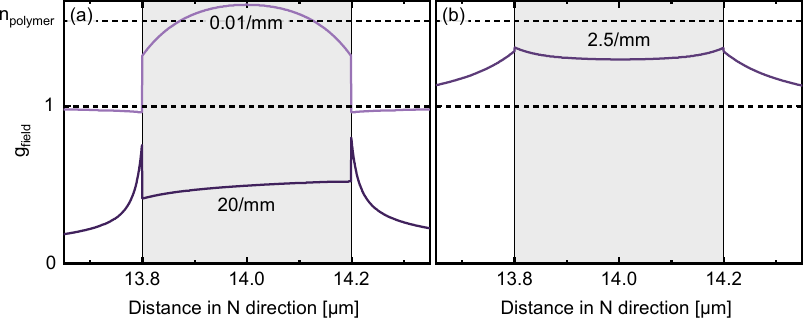}
	\caption[]{Superchiral fields in helicoidal waveguides. $g_\mathrm{field}$ was evaluated for the LCP fundamental mode along the N direction through the center of the waveguide. This is an extension of Fig.\ 7
	(b) of the main text. (a) At low twist rates (0.01/mm), $g_\mathrm{field}$ is close to $n$ like in circularly polarized plane wave. At high twist rates (20/mm) the field becomes linearly polarized resulting in an overall lower value of $g_\mathrm{field}$. (b) At an intermediate twist rate (2.5/mm) the field becomes superchiral on the surface of the waveguide ($g_\mathrm{field}>1$).}	
	\label{fgr:gfield_LineCuts_SI}
\end{figure}

In typical chiral sensing experiments of molecules, the quantity of interest is the difference in absorbance between right- and left-circularly polarized light normalized to the average absorbance. This quantity is called Kuhn's dissymmetry factor or $g$-factor \cite{Barron2004}:
\begin{equation}
	g = \frac{A^R-A^L}{\frac{1}{2} \left(A^R+A^L \right)}.
\end{equation}
Approximating the small molecule as a dipole, a relationship between the optical chirality density of the electromagnetic field and the measured response of the chiral molecule was shown theoretically \cite{Tang2010} and experimentally \cite{Tang2011} to be:
\begin{equation} \label{gfactorfield}
	g = g_\mathrm{molecule} \cdot \underbrace{ \frac{cC}{2 \langle U_e \rangle \omega}}_{g_\mathrm{field}},
\end{equation}
where $ \langle U_e \rangle = \frac{1}{2} \langle \vb{D} \cdot \vb{E} \rangle = \frac{1}{4} \abs{\vb{D}} \abs{\vb{E}} $ is the time-averaged energy density of the electric field. $g_\mathrm{molecule}$ is the $g$-factor that would be obtained in a measurement with plane wave CPL and only depends on the properties of the molecule (energy levels and transition moments). The second factor $g_\mathrm{field}$ only depends on the properties of the field.
\\ \\
Eq.\ (\ref{gfactorfield}) shows that the measured $g$-factor of a given molecule can be enhanced by engineering a suitable field with a large value of $g_\mathrm{field}$, which is referred to as superchiral field. This can be achieved by reducing the amplitude of the electric field while maintaining a high chirality density (e.g., strong magnetic field, and circular polarization of $\vb{E}$ and $\vb{H}$). Throughout this work, the quantity $g_\mathrm{field}$ is used to characterize the chirality of the fields in the twisted waveguides. For linear media one can simplify $g_\mathrm{field}$ as:
\begin{equation}
	g_\mathrm{field} = Z_0 \frac{\Im{\vb{E} \cdot \vb{H^*}}}{\abs{\vb{E}}^2}, 
\end{equation}
with $Z_0 = \sqrt{\frac{\mu_0}{\epsilon_0}}$ being the vacuum impedance. For a linearly polarized wave $g_\mathrm{field}$ is equal to 0 and plugging in the values for a circularly polarized plane wave from Eq.\ (\ref{CirPol}) yields:
\begin{equation}	
	g_\mathrm{field} = s \, n \hspace{3 mm} (\mathrm{circularly \ polarized \ plane \ wave}).
\end{equation}
Consequently, any field distribution with $\abs{g_\mathrm{field}} > n$ can increase the measured $g$-factor and is referred to as superchiral field. As shown in Fig.\ 7
(b) of the main text such superchiral fields can occur on the surface of helicoidal waveguides. Concerning this figure, additional line cuts of $g_\mathrm{field}$ along the N direction for different twist rates can be found here in \ref{fgr:gfield_LineCuts_SI}.

\section{Photonic spin Hall and orbital Hall effect in helical waveguides} \label{SI_SHE}

Circularly polarized light traveling along curved trajectories has been observed to split into two beams of opposite spin \cite{Bliokh2008}. This unexpected behavior is an example of the photonic spin Hall effect, similar to the electronic spin Hall effect observed in semiconductors \cite{Kato2004, Wunderlich2005}. \\

The photonic spin Hall effect is based on the spin-orbit interaction of photons, which can be derived by approximating the complex electromagnetic fields as a single wavepacket, describing only the motion of its center of gravity \cite{Liberman1992, Bliokh2015}. Such a "semi-geometrical approximation" can be applied when the trajectory of the wavepacket changes on lengthscales much larger than the wavelength. More specifically the parameter $\mu = \lambda_0/[2 \pi \, \mathrm{min}(\bar{\kappa}^{-1}, \tau^{-1})]$ should be small. For a helical path, $\bar{\kappa}^{-1}$ and $\tau^{-1}$ are always larger than the radius $\rho$ of the helix. Therefore, $\mu < 9 \times 10^{-3}$ with the parameters used in this work ($\rho = 14 \ \upmu \rm m, \; \lambda_0 = 770 \ \mathrm{nm}$) and the approximation is principally valid. \\

The motion of the wavepacket can then be described to first order in $\mu$ as \cite{Bliokh2004a, Bliokh2008}:
\begin{equation} \label{eq:SpinHall}
	\frac{\partial \vb{\hat{T}}}{\partial s} = \bar{\kappa} \, \vb{\hat{N}}, \hspace{8mm}
	\frac{\partial \vb{r}}{\partial s} = \vb{\hat{T}}  \underbrace{- \hat{S}_3 \, \frac{\lambda_0}{2 \pi n} \, \bar{\kappa} \, \vb{\hat{B}}}_{\mathrm{Spin{\text -}orbit \ splitting}},
\end{equation}
where $\vb{r}$ describes the position of the wavepacket parametrized by the arc length $s$ of its trajectory and ($\vb{\hat{T}}$, $\vb{\hat{N}}$, $\vb{\hat{B}}$) are the unit vectors of the Frenet-Serret frame for the trajectory that the light would take without spin-orbit coupling. The Frenet-Serret coordinate system is defined in Sec.\ \ref{FrenetSerretSec}. $\bar{\kappa}$ is the local curvature, $\hat{S}_3$ is the third Stokes parameter and $n$ is the refractive index. \\

The term on the right side is a consequence of the spin-orbit interaction of photonic wavepackets and results in a spin-dependent splitting of the beam along the B axis. For the helical waveguide with the highest investigated twist rate ($\rho = 14 \ \upmu \rm m, \; P=50  \ \upmu \rm m, \; \lambda_0 = 770 \ \mathrm{nm}, \; n=1.54$), the magnitude of the spin-orbit splitting term is equal to $\frac{\lambda_0}{2 \pi n} \, \bar{\kappa} = 4.3 \times 10^{-3}$. This value is about three orders of magnitude larger than in a well-known free-space demonstration of the photonic spin Hall effect. In this demonstration, a beam was constrained to a helical path at the inner surface of a macroscopic glass cylinder ($\rho = 8 \ \mathrm{mm}, P=16 \ \mathrm{mm}, \; \lambda_0 = 633 \ \mathrm{nm}, \; n \approx 1.5, \; \frac{\lambda_0}{2 \pi n} \, \bar{\kappa} = 7.6 \times 10^{-6}$) \cite{Bliokh2008}. In this case, the LCP and RCP beams are found to be displaced along the $z$ axis (i.e., along the twist axis), which follows from integration of the spin-orbit splitting term over one turn of the helix. In other words, the two beams propagate along helices with the same radius but different pitch lengths. Experimentally, a splitting of about $2 \ \upmu \rm m$ was observed after a propagation distance of 96 mm. \\

The situation is different for a twisted waveguide because its LCP and RCP eigenmodes have the same pitch length as the confining structure and do not change as they propagate along the waveguide. Therefore, a splitting along the axial direction as in Ref. \cite{Bliokh2008} cannot occur. Instead, we observe a spin-dependent splitting along the N direction for the Frenet-Serret waveguide. The splitting occurs in the T component of the Poynting vector for its multimode variant (cf. Fig.\ 6
(b) of the main text) and in the transverse component of the Poynting vector for its single-mode variant (cf. Fig.\ 7
(a) of the main text). A shift in B direction was not found in the simulations. \\

We do not attempt to provide a complete description of the photonic spin Hall effect in helical waveguides, which would need to take the confinement provided by the waveguide into account in addition to the spin-orbit splitting in Eq.\ (\ref{eq:SpinHall}). However, the fact that the photonic spin Hall effect causes our observed splitting can be made plausible by the following thought: Assuming that the confinement of the waveguide is not spin-dependent, an RCP and an LCP field with the same center of gravity experience an identical confining "force" (since they are located at the same position) but opposite spin-orbit shifts. Therefore, these fields would separate during infinitesimal propagation and cannot be stable eigenmodes. If their centers of gravity are different, on the other hand, they can experience different confining "forces" which can compensate the opposing spin-orbit shifts yielding stable eigenmodes. We found empirically that this separation occurs along the N direction. \\

Akin to the spin-orbit interaction, there is also an orbit-orbit interaction of photons, which couples the intrinsic OAM of a mode (caused by an $\mathrm{e}^{\mathrm{i} l \phi}$ phase dependence) with its orbital motion (extrinsic OAM). The OAM-dependent spatial shifts caused by this interaction are referred to as photonic orbital Hall effect \cite{Fedoseyev2001, Bliokh2006a}. The equations of motion for $\mu \ll 1$ are analogous to those of the spin Hall effect with the spin being replaced by the total angular momentum \cite{Bliokh2006a}:
\begin{equation}
	\frac{\partial \vb{\hat{T}}}{\partial s} = \bar{\kappa} \, \vb{\hat{N}}, \hspace{8mm}
	\frac{\partial \vb{r}}{\partial s} = \vb{\hat{T}} - (s+l) \, \frac{\lambda_0}{2 \pi n} \, \bar{\kappa} \, \vb{\hat{B}}.
\end{equation}
Therefore, the orbit-orbit interaction results in the same spatial splitting as the spin-orbit interaction but with a potentially much larger amplitude since the value of $l$ can become arbitrarily large (within the number of available spatial modes). The orbit-orbit interaction might explain the OAM-dependent splittings in the intensity distribution of the modes shown in Fig.\ 4
(c,d) of the main text. This splitting can already be observed at a low twist rate of 5/mm while splittings in the fundamental modes only become apparent at higher twist rates of 10/mm - 20/mm, in line with the different magnitudes of the spin-orbit and orbit-orbit interaction.

\newpage
\section{Modal properties of all investigated waveguides} \label{SI_AllPlots}

To allow a better comparison of the modal properties across different waveguide types and twist rates, detailed plots are available on the next pages. The following optical properties (defined in Sec.\ \ref{SI_DefsOptQuantities}) were evaluated:

\begin{itemize}
	\item Longitudinal component of the Poynting vector $S_\mathrm{T}$.
	\begin{itemize}
		\item Corresponds to the intensity that would be measured at the output of the waveguide.
	\end{itemize}
	\item Transverse component of the Poynting vector $S_{\mathrm{NB}}$. 
	\begin{itemize}
		\item Both, the longitudinal and transverse components were scaled by the same value so that their magnitude can be compared.
		\item If this component curls around the center of the waveguide in the azimuthal direction, the mode contains OAM. A small fraction of OAM is present in the fundamental modes because spin and OAM are only well separated as long as the paraxial approximation is strictly valid \cite{Aiello2015}. For the single-mode waveguides in particular, the paraxial approximation is not well satisfied and some of the spin angular momentum is converted to OAM while the total angular momentum is conserved \cite{Bliokh2015}.
	\end{itemize}
	\item Transverse component of electric and magnetic field, $\vb{E}_{\mathrm{NB}}$ and $\vb{H}_{\mathrm{NB}}$, respectively.
	\begin{itemize}
		\item White ellipses denote shape and orientation of the polarization ellipse within the NB plane (i.e., neglecting the longitudinal field component). Arrows indicate the direction of rotation of the polarization vector in time.
	\end{itemize}
	\item Spin vector of electric and magnetic field, $\vb{s}_{\vb{E}}$ and $\vb{s}_{\vb{H}}$, respectively.
	\begin{itemize}
		\item The magnitude of the spin vector corresponds to the Stokes parameter $\hat{S}_3$ of the respective field.
		\item The transverse component of the spin vector is shown on top of $\hat{S}_3$ in the square-shaped plots.
		\item The longitudinal component of the spin vector can be seen in the projection of the spin vector onto the TB plane shown in the rectangular box below. The height of the box corresponds to the maximal value of 1. The scaling of the arrows in the box and the rectangle is the same.
	\end{itemize}
	\item Phase of one of the transverse components of the electric and magnetic field.
	\begin{itemize}
		\item The phase of the B component is shown unless the N component is dominant. Labels in the plots indicate which component is shown.
		\item The absolute value of the phase varies between plots.
	\end{itemize}
	\item Characterization of the chirality of the fields in the form of $g_{\mathrm{field}}$.
	\begin{itemize}
		\item If $\abs{g_{\mathrm{field}}}>n$, the field is superchiral.
		\item $g_{\mathrm{field}}$ is not shown for multimode waveguides since its magnitude is close to $n$ at all twist rates. 
	\end{itemize}
\end{itemize}
\newpage
\begin{figure}[H]
	\centering
	\includegraphics[scale=1]{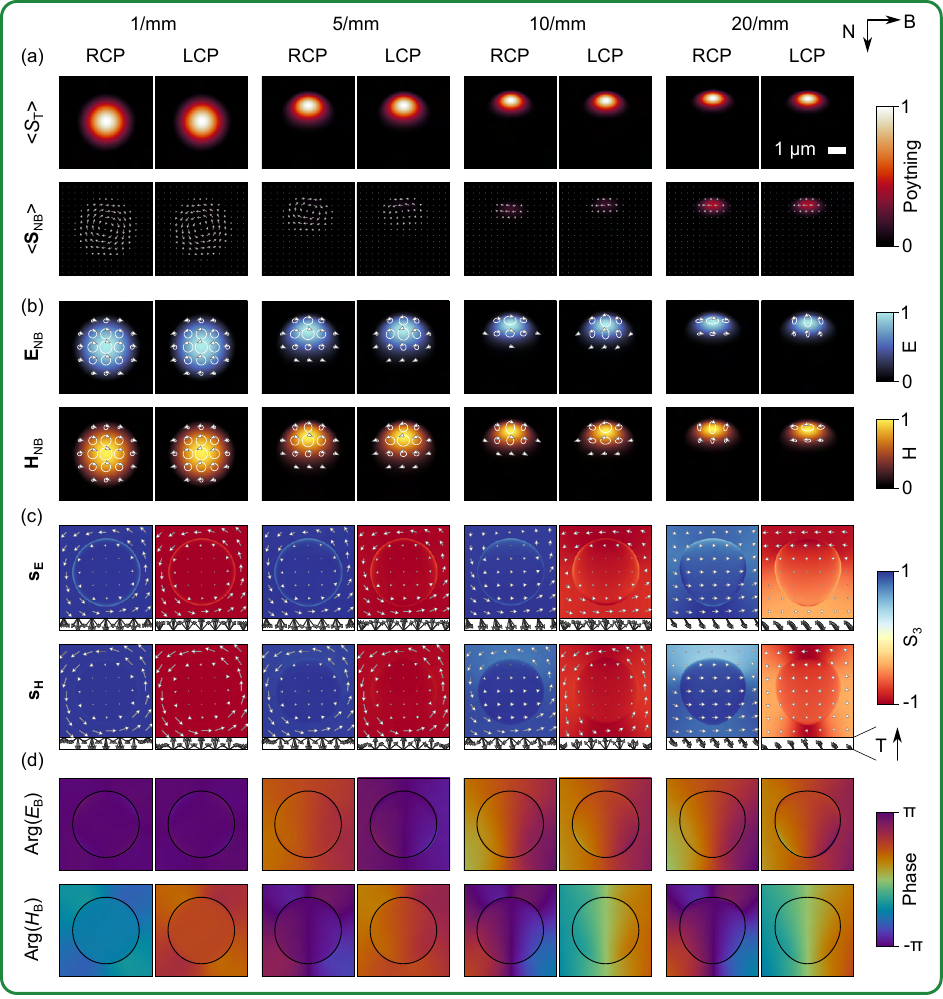}
	\caption[]{Spatial properties of fundamental modes in the multimode Frenet-Serret waveguide ($V=17.25$) depicted in the TNB frame. Four different twist rates from 1/mm to 20/mm are shown. Modes remain circularly polarized up to the highest twist rate. The cross sections of the waveguides are shown as black lines in (d). Since the cross section is approximated as an ellipse in the $xy$ plane (cf. Sec.\ \ref{SI_CrossSecFormula}), the cross section in the TNB frame is not perfectly circular. Quantities depicted in (a-d) are explained at the beginning of this chapter.}	
	\label{fgr:Multimode_Frenet_SI}
\end{figure}

\begin{figure}[H]
	\centering
	\includegraphics[scale=1]{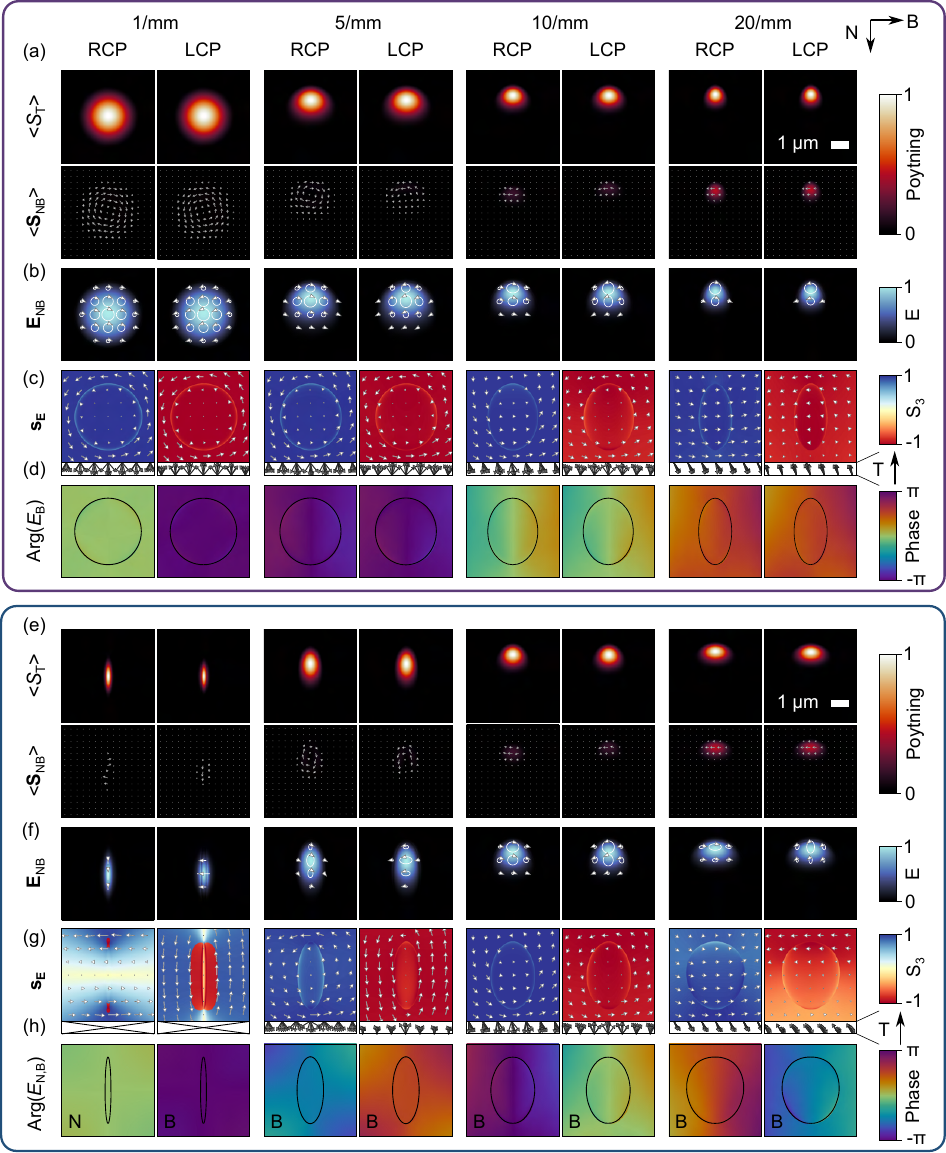}
	\caption[]{Spatial properties of fundamental modes in the multimode helicoidal waveguide (a-d) and the multimode Overfelt waveguide (e-h) depicted in the TNB frame. Four different twist rates from 1/mm to 20/mm are shown. The cross sections of the waveguides are shown as black lines in (d,h). Longitudinal components of the spin vector in the first two entries in (g) are not shown because they are negligibly small. Quantities depicted in (a-h) are explained at the beginning of this chapter.}	
	\label{fgr:Multimode_Helicoidal+Overfelt_SI}
\end{figure}

\begin{figure}[H]
	\centering
	\includegraphics[scale=1]{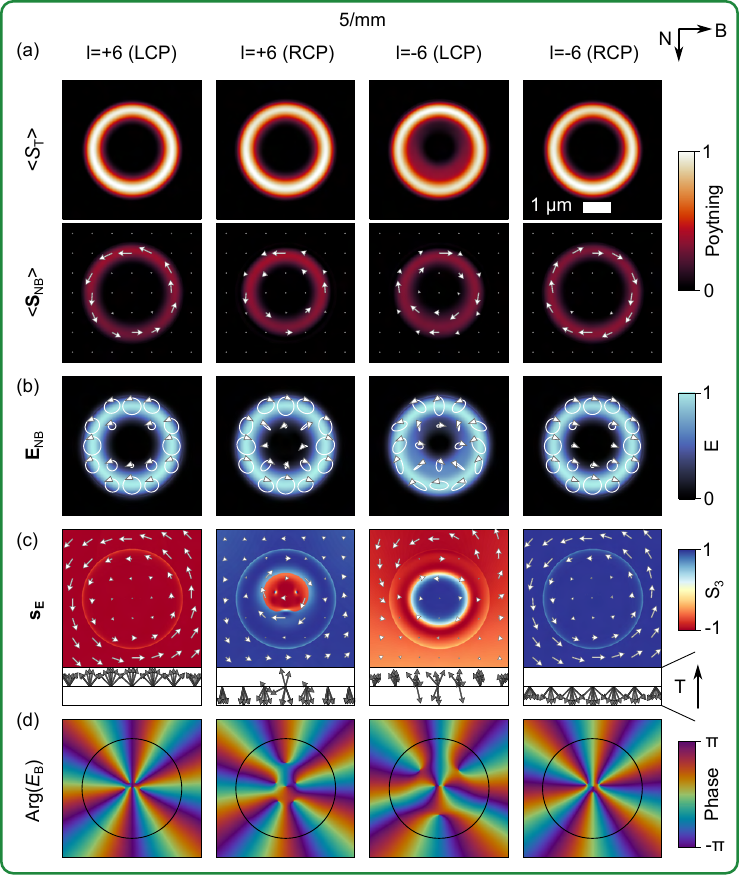}
	\caption[]{Spatial properties of OAM modes ($\abs{l}=6$) in the multimode Frenet-Serret waveguide ($V=17.25$) depicted in the TNB frame. Modes are evaluated at an intermediate twist rate of 5/mm. The cross sections of the waveguides are shown as black lines in (d). The two modes in the center [$l=+6$ (RCP), $l=-6$ (LCP)] are coupled to higher-order modes causing the changes in their phase and polarization properties. Quantities depicted in (a-d) are explained at the beginning of this chapter.}	
	\label{fgr:Multimode_FrenetOAM_SI}
\end{figure}

\begin{figure}[H]
	\centering
	\includegraphics[scale=1]{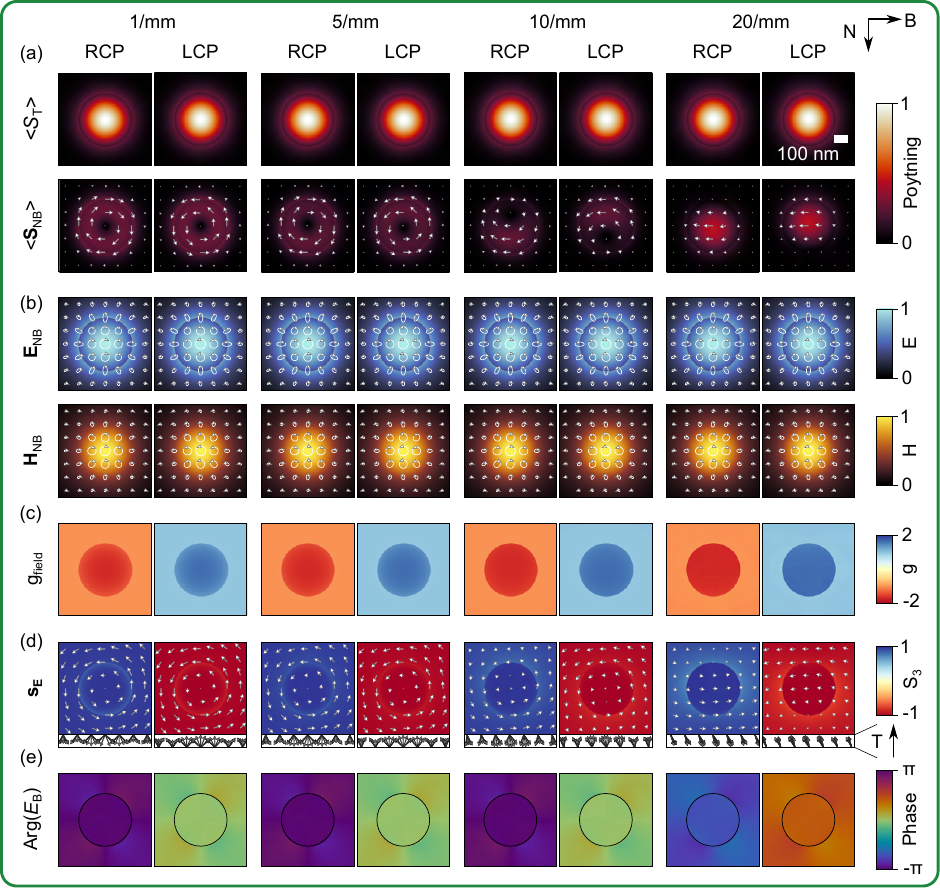}
	\caption[]{Spatial properties of modes in the single-mode Frenet-Serret waveguide ($V=1.92$) depicted in the TNB frame. Four different twist rates from 1/mm to 20/mm are shown. Modes remain circularly polarized up to the highest twist rate. The cross sections of the waveguides are shown as black lines in (e). Quantities depicted in (a-e) are explained at the beginning of this chapter.}	
	\label{fgr:Singlemode_Frenet_SI}
\end{figure}

\begin{figure}[H]
	\centering
	\includegraphics[scale=1]{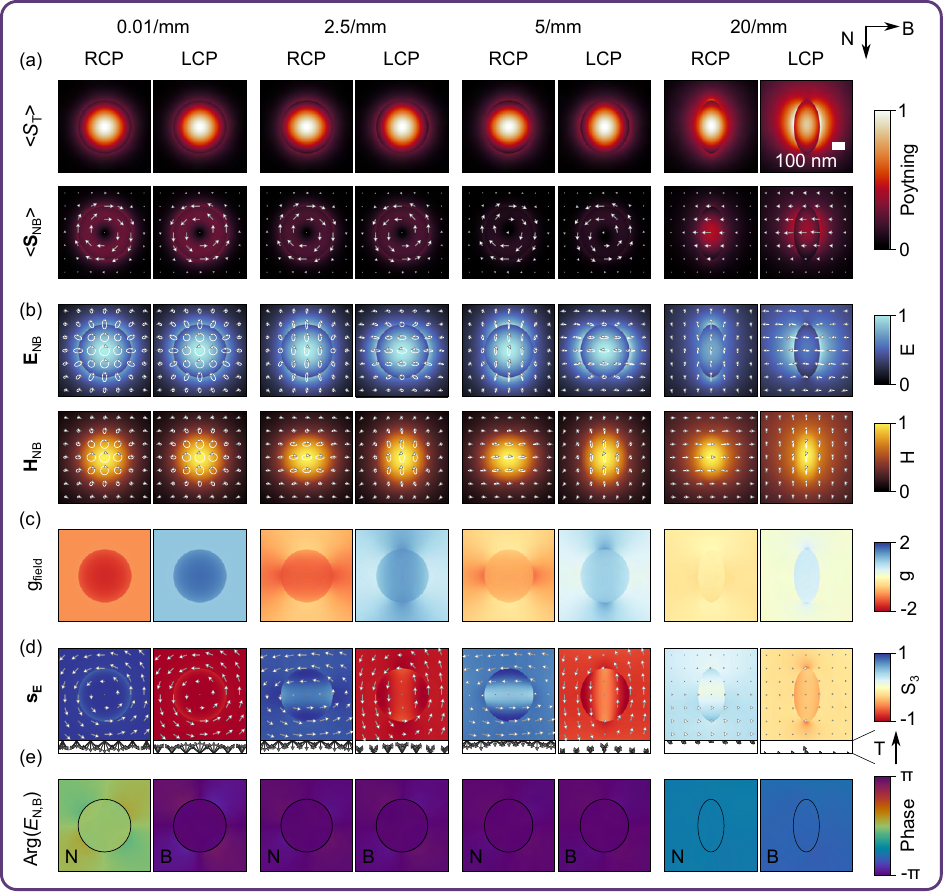}
	\caption[]{Spatial properties of the  modes in the single-mode helicoidal waveguide depicted in the TNB frame. Twist rates from 0.01/mm to 20/mm were selected here to better show the chiral properties of the mode (c). The cross sections of the waveguides are shown as black lines in (e). Quantities depicted in (a-e) are explained at the beginning of this chapter.}	
	\label{fgr:Singlemode_Helicoidal_SI}
\end{figure}

\begin{figure}[H]
	\centering
	\includegraphics[scale=1]{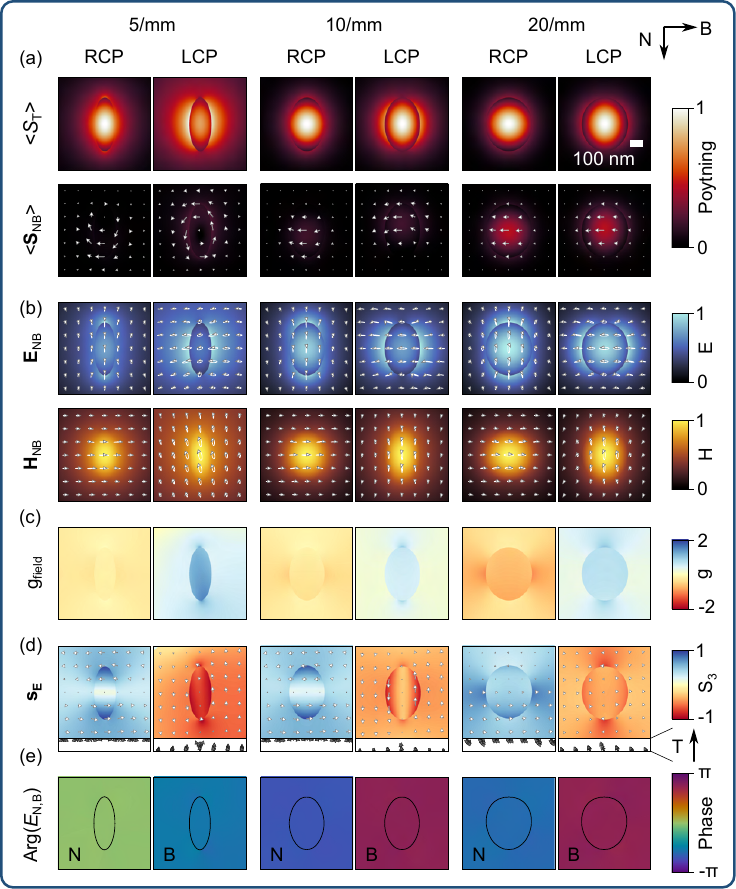}
	\caption[]{Spatial properties of the  modes in the single-mode Overfelt waveguide depicted in the TNB frame. Only three different twist rates from 5/mm to 20/mm are shown here. At lower twist rates the cross section becomes very narrow resulting in high bend losses that cannot be handled reliably by the solver. The cross sections of the waveguides are shown as black lines in (e). Quantities depicted in (a-e) are explained at the beginning of this chapter.}	
	\label{fgr:Singlemode_Overfelt_SI}
\end{figure}

\bibliography{1_Bib}